\documentclass{article}

\usepackage{graphicx}
\usepackage{graphicx,xcolor}
\usepackage{subcaption}
\usepackage{multicol,multirow}
\usepackage{amsmath,amssymb,amsfonts}
\usepackage{mathrsfs}
\usepackage{amsthm}
\usepackage{rotating}
\usepackage{appendix}
\usepackage{authblk}
\usepackage{amsthm,amssymb,bm}
\usepackage{mathtools}
\usepackage{verbatim}
\usepackage{fullpage, fourier, charter}

\usepackage[colorlinks,allcolors=magenta]{hyperref}

\theoremstyle{definition}

\numberwithin{equation}{section}
\def\bse{\begin{subequations}}
\def\ese{\end{subequations}}

\title{A regularized continuum model for traveling waves and dispersive shocks of the granular chain}
\author[1]{Su Yang}
\author[2]{Gino Biondini}
\author[3]{Christopher Chong}
\author[1]{Panayotis G.~Kevrekidis}

\affil[1]{Department of Mathematics and Statistics, University of Massachusetts, Amherst, 01003-4515, Massachusetts, USA}
\affil[2]{Department of Mathematics, University at Buffalo, Buffalo, NY 14260-2900, USA}
\affil[3]{Department of Mathematics, Bowdoin College, Brunswick, ME 04011, USA}

\date{}

\begin{document}

\maketitle

\abstract{In this paper we focus on a discrete physical model describing granular crystals, whose equations of motion can be described by a system of differential difference equations (DDEs). After revisiting earlier continuum approximations, we propose a regularized continuum model variant to approximate the discrete granular crystal model through a suitable partial differential equation (PDE). We then compute, both analytically and numerically, its traveling wave and periodic traveling wave solutions, in addition to its conservation laws. Next, using the periodic solutions, we describe quantitatively various features of the dispersive shock wave (DSW) by applying Whitham modulation theory and the DSW fitting method. Finally, we perform several sets of systematic numerical simulations to compare the corresponding DSW results with the theoretical predictions and illustrate that the continuum model provides a good approximation of the underlying discrete one.}

\section[Introduction]{Introduction}

Dispersive shock waves, which are coherent, non-stationary, multiscale nonlinear wave structures that connect states of different amplitude 
via an expanding modulated wave train, have gained significant attention
recently in a variety of settings, including ultracold gases, optics, superfluids, electron beams, and plasmas 
\cite{scholar,Mark2016,Whitham74}. 
In particular, dispersive shock waves in one-dimensional (1D) nonlinear lattices (to be called lattice DSWs here) have been explored numerically, and even experimentally, in several works \cite{first_DSW,Nester2001,Hascoet2000,Herbold07,shock_trans_granular,Molinari2009,HEC_DSW}. 
Much of the motivation for the above studies stems from granular chains, which consist of closely 
packed arrays of particles that interact elastically upon compression. They 
have received much recent attention due to their potential in applications,
recent advances in experimental platforms 
and the mathematical richness of the underlying equations
and of the waveforms that arise therein. 
We refer the reader to Refs.~\cite{Nester2001,granularBook,yuli_book,gc_review,sen08} for comprehensive reviews
on the subject of granular chains. 
While this article focuses on DSWs in granular chains, 
lattice DSWs, in general, are of broad physical interest, as similar
structures have been experimentally observed, e.g., in nonlinear
optics of waveguide arrays~\cite{fleischer2}.
It is important to also highlight in this context 
another setup that has recently emerged, namely tunable magnetic
lattices~\cite{talcohen}. Here, ultraslow shock waves can arise and have 
been experimentally imaged offering yet another 
platform where such patterns can be visualized
in their space-time evolution.

Beyond direct numerical simulations, lattice DSWs have been studied through a variety of lenses. 
The classical approach, based on Whitham modulation theory, was explored in works like
\cite{Venakides99,DHM06,blochkodama,physd2024v469p134315}.  In general lattice
settings, however, the modulation equations corresponding to lattice DSWs can be quite cumbersome, and thus other approaches are also desirable. 
Examples include analytical techniques to estimate the leading and trailing amplitudes
\cite{marchant2012}, the DSW fitting method~\cite{El2005} applied to discrete settings~\cite{Sprenger2024}, reduction
of dynamics to a planar ODE (possibly in 
a data-driven manner)~\cite{CHONG2022} and integrable approximations \cite{Ari2024}.

In the present work we derive and examine an analytically tractable continuum model that does not rely on assumptions of small amplitude,
and we use the results to quantitatively describe the DSWs of the granular chain.
Specifically, the outline of this work is the following.
In section~\ref{s:models} we introduce the theoretical setup and derive the continuum model.
In sections~\ref{s:solitarywaves} and~\ref{s:periodicwaves} 
we obtain analytical expressions for the solitary waves and periodic traveling waves (respectively) of the continuum model.
In section~\ref{s:conservationlaws} we present the conservation laws of the model and 
in section~\ref{s:whitham} we use all of the above ingredients to formulate the corresponding Whitham modulation theory.
In section~\ref{s:riemann} we derive and study the harmonic limits and soliton limits of the modulation equations where explicit results can
be obtained, 
and  we use these reductions to study Riemann problems and characterize the limiting features of the corresponding DSWs.
Finally, in section~\ref{s:numerics} we validate the results by comparison with systematic numerical simulations.
Section~\ref{s:conclusions} ends this work with some concluding remarks and some future challenges 
along the emerging direction of discrete 
dispersive hydrodynamics.

\section{Models and theoretical setup}\label{s:models}

In this work we focus on granular lattice dynamical systems. 

The non-dimensional equations of motion
(where the constant elastic and geometric 
prefactors have been absorbed
through a suitable rescaling) is given by the following differential-difference equations (DDEs)~\cite{Nester2001,granularBook,yasuda}:
\begin{equation}
    \ddot u_{n} = \left[\delta_{0} + u_{n-1} -u_{n}\right]^{p}_{+} - \left[\delta_{0} +u_{n}-u_{n+1}\right]^{p}_{+},
\label{Displacement ODE}
\end{equation}
where $\delta_{0}$ denotes the precompression constant, $u_{n}$ the displacement of the $n$th particle from its equilibrium position, and $\left[f\right]_{+} = \text{max}(f,0)$, models the fact that there is no
force when the particles come out of contact. 
The case of $p=3/2$ corresponds to a lattice of spherical particles. It is relevant to mention in passing that in other settings such as O-rings, cylindrical particles
or hollow spheres the nature of the force (and hence exponent) may vary~\cite{Johnson} and hence we will
maintain $p$ as a general parameter in our
considerations herein.
Equation~\eqref{Displacement ODE} possesses traveling wave solutions~\cite{Nester2001,sen08,granularBook,yuli_book}, 
even though their exact form is not known 
analytically (but for relevant approximations see e.g.,~\cite{Yuli2010}).
Moreover, numerical investigations of Eq.~\eqref{Displacement ODE} reveal the formation of dispersive shocks in certain regimes \cite{Herbold07,shock_trans_granular,Molinari2009,yasuda}.

In a recent work, we studied the DSWs produced by Eq.~\eqref{Displacement ODE} when $\delta_0\ne0$,
making use of two different approximations for it:
the Toda lattice and the Korteweg-de\,Vries equation \cite{Ari2024}.
In the present work, we focus on the case where $\delta_{0}=0$, in which case the model does not admit a meaningful linear limit in the case of zero
background.

It will be convenient to work with the  strain variable $r_{n} =  u_{n-1} - u_{n}$, which has a physically meaningful interpretation in terms of granular
crystals (the amount of compression between adjacent particles),
and is common when describing granular crystals, see, e.g., the authoritative book of~\cite{Nester2001}.
We note in passing, however, that some authors also use the variable $u_{n} - u_{n-1}$ 
for the relevant analysis \cite{DHM06}.
This allows one to rewrite Eq.~\eqref{Displacement ODE} as
\begin{equation}
    \ddot r_{n} = \left(r_{n+1}\right)^{p} - 2\left(r_{n}\right)^{p} + \left(r_{n-1}\right)^{p},
\label{e:Strain ODE}
\end{equation}
where we dropped (here and henceforth) the subscript $+$ under the assumption that the strains are non-negative.
The linearized dispersion relation 
for small-amplitude wave solutions of the form
$r_n(t) = A + B e^{i(kn - \omega t)}$, with $|B/A| \ll 1$,
is given by
\begin{equation}
    \omega^2 = 4pA^{p-1} \sin^2\left(\frac{k}{2}\right)\,,
\label{e:disp}
\end{equation}
where $k$ and $\omega$ are the wavenumber and frequency, respectively.

Conversely, dispersive shock waves, which are one the main concerns of this paper, arise in Eq.~\eqref{e:Strain ODE} when initialized with Riemann 
initial data
\begin{equation}\label{step}
  r_n(0) = \begin{cases}
             r_-, \quad & n \leq 0 \\
             r_+, \quad & n > 0 ,
           \end{cases} \qquad
     \dot{r}_n(0) = \begin{cases}
             v_-, \quad & n \leq 0 \\
             v_+, \quad & n > 0 .
           \end{cases} \qquad        
\end{equation}

We wish to use a dispersive long-wavelength model to approximate \eqref{e:Strain ODE}. 
To this end, we introduce an associated smallness parameter $0 < \epsilon \ll 1$ and the following slowly varying spatial and temporal variables
\begin{equation}
    X = \epsilon n, \hspace{3mm} T = \epsilon t.
\end{equation}
Using the ansatz $r_{n}(t) = r(X,T)$ and substituting into Eq.~\eqref{e:Strain ODE} leads to
\begin{equation}
    \epsilon^{2}r_{TT} = r^{p}(X+\epsilon,T) + r^{p}(X-\epsilon,T) - 2r^{p}(X,T).
    \label{Transformed strain equation}
\end{equation}
A Taylor expansion of Eq.~\eqref{Transformed strain equation} then yields, to leading order,
the PDE
\begin{equation}
    r_{TT} = \left(r^{p}\right)_{XX}.
\label{e:PDEleadingorder}
\end{equation}
Equation~\eqref{e:PDEleadingorder} was already studied in \cite{yasuda} (see also its earlier derivation
in~\cite{dcdsa}),
where it was shown to correctly capture the wave breaking in the solutions of Eq.~\eqref{e:Strain ODE}.
However, Eq.~\eqref{e:PDEleadingorder} is a non-dispersive model, and as such it does not give rise to the formation of DSWs.
Indeed, looking for small-amplitude plane-wave solutions of Eq.~\eqref{e:PDEleadingorder} in the form $r(X,T) = A + B \,e^{i(K X - \Omega T)}$, where $|B/A|\ll1$ and $K$ and $\Omega$ are the
wavenumber and frequency with respect to the variables $X,T$,
yields the linearized dispersion relation $\Omega^2 = pA^{p-1}\,K^2$, for which $d^2\Omega/dK^2\equiv0$
(hence the PDE non-dispersive).
Note that we can relate the wavenumber and frequency of the PDE model to the original lattice variables
through the relationship 
\begin{equation} \label{eq:Kk}
  K= \epsilon^{-1} k, \qquad \Omega= \epsilon^{-1} \omega  
\end{equation}

In order to be able to describe DSWs, 
we include next order in the Taylor expansion of Eq.~\eqref{Transformed strain equation}
by keeping terms up to $\mathcal{O}\left(\epsilon^{2}\right)$.
Doing so yields the PDE
\begin{equation}
    r_{TT} = \left(r^{p}\right)_{XX} + \frac{\epsilon^{2}}{12}\left(r^{p}\right)_{XXXX}.
\label{not good model}
\end{equation}
This model, originally proposed in~\cite{Ahnert_2009},
is ill-posed, however, due to large wavenumber instabilities. 
Indeed, looking for plane-wave solutions as before yields 
$\Omega^2 = p A^{p-1} K^2 ( 1 - \frac1{12} \epsilon^2 K^2)$.
Note that $\Omega$ is purely imaginary for sufficiently large wavenumbers 
(i.e., $|K| > 2\sqrt{3}/\epsilon$). 
Thus, small wavelength oscillations are unstable.
Moreover, the imaginary part of $\Omega$ is unbounded. 
Therefore, it is expected that Eq.~\eqref{not good model} is ill-posed as an initial-value problem.

\begin{figure}[t!]
    \centering
    \includegraphics[width=0.5\linewidth]{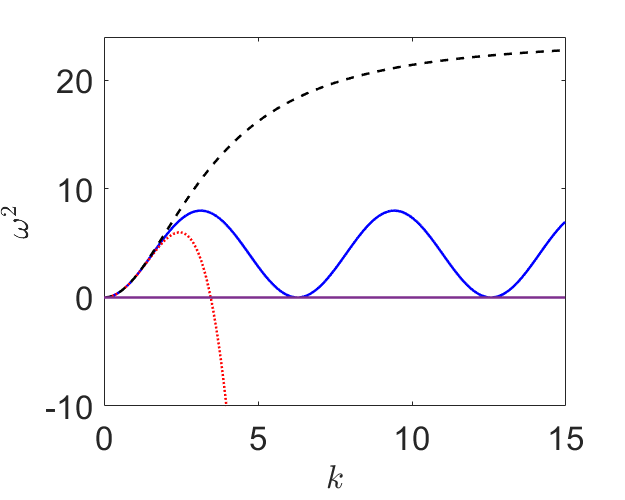}
    \caption{Comparison of the linearized dispersion relations of different models in terms of $(k,\omega)$. The solid blue line corresponds to the lattice dispersion relation~\eqref{e:Strain ODE}, with Brillouin zone given by $[0,\pi]$. We show larger wavenumbers for comparison purposes. The red dotted curve is the dispersion relation of~\eqref{not good model} (and~\eqref{Pikovsky's continuum model}, which happens to be identical). For $k>2 \sqrt{3}$ we see that $\omega^2 < 0$, which leads to {ultraviolet (i.e. high wavenumber) instability and ultimately ill-posedness}.  The black dashed curve depicts the linearized dispersion relation of regularized model~\eqref{Target PDE model}, which has a horizontal asymptote
    ($\omega^2 \rightarrow 24$ {for the specific values used here}).
    {For all curves,} the parameters are chosen to be $p = 2, A = 1$.
    }
\label{fig:dispersionrelations}
\end{figure}

On the other hand, Eq.~\eqref{not good model} can be regularized in a straightforward way,
following \cite{rosenau1,rosenau2}. 
In particular,
since
\begin{equation}
    r_{TT} = \left(r^{p}\right)_{XX} + \frac{\epsilon^{2}}{12}\left(r^{p}\right)_{XXXX} 
    = \left(r^{p}\right)_{XX} + \mathcal{O}(\epsilon^{2}),
    \label{Rewrite (5)}
\end{equation}
we have that
\begin{equation}
    \left(r^{p}\right)_{XXXX} = \left(\left(r^{p}\right)_{XX}\right)_{XX} 
    = \left(r_{TT} - \mathcal{O}(\epsilon^{2})\right)_{XX} 
    = r_{TTXX} + \mathcal{O}(\epsilon^{2}).
    \label{first substitution}
\end{equation}
The substitution of \eqref{first substitution} into \eqref{Rewrite (5)} then yields, up to $\mathcal{O}\left(\epsilon^{4}\right)$ terms
\begin{equation}
    r_{TT} - \frac{\epsilon^{2}}{12}r_{XXTT} = \left(r^{p}\right)_{XX}.
    \label{Target PDE model}
\end{equation}
If we now look for plane wave solutions of~\eqref{Target PDE model}, we obtain
\begin{equation}
\Omega^2 = \frac{pA^{p-1}K^2}{\left(1 + \frac{\epsilon^2}{12}K^2\right)}\,, 
\label{e:NLWPDEO2disprel}
\end{equation}
which is a perfectly well-behaved dispersion relation, real-valued for all $K\in\mathbb{R}$.
Figure~\ref{fig:dispersionrelations} shows Eq.~\eqref{e:NLWPDEO2disprel} compared to 
Eq.~\eqref{e:disp} and other relevant approximations (see below).  Note that
the PDE dispersion relations are independent of $\epsilon$ if expressing in terms of the original lattice variables (using Eq.~\eqref{eq:Kk}).

An additional interesting observation here is that
even when the model is nonlinearly dispersive
with $p \neq 1$, the additional dispersive 
effects of $\mathcal{O}\left(\epsilon^{2}\right)$
appear at the linear level herein. This is
distinct from the (ill-posed)
previously proposed continuum approximations
such as those of~\cite{Ahnert_2009}, as well
as that of~\cite{Nester2001}, where the continuum
limit is taken at the level of displacements,
rather than that of strains. We will compare
these different models further, e.g., at the level
of their traveling waves in the next section.

Equation~\eqref{Target PDE model} is the primary model of interest in this work. 
In particular, we will show that Eq.~\eqref{Target PDE model} accurately captures, both qualitatively and quantitatively, the solitary waves, periodic traveling waves and DSWs of Eq.~\eqref{Displacement ODE}. 
Recall that Zabusky and Kruskal derived the Korteweg-de-Vries (KdV) equation with small dispersion as a long-wave approximation of the Fermi-Pasta-Ulam-Tsingou (FPUT) problem~\cite{Zabusky,FPUreview}.
\textit{We claim that \eqref{Target PDE model} plays for the granular lattice model~\eqref{e:Strain ODE} a similar role that the KdV equation plays for the FPUT problem}.
Specifically, the claim is that, just like the KdV equation provided a useful continuum model
to describe, at least in the long-wave limit, the dynamics of solutions of the FPUT problem,
so does the PDE~\eqref{Target PDE model} for the granular lattice model~\eqref{e:Strain ODE}.

\section{Solitary wave solutions}\label{s:solitarywaves}

While the ultimate goal of this work is to characterize the DSWs of the granular chain, to do so we must first review its solitary waves and periodic traveling wave solutions. We delve into the analysis of these
states in this and the next section.

\subsection{Derivation of solitary waves}

Our first goal is to find the traveling solitary wave corresponding to the model \eqref{Target PDE model}. 
To this end, we first introduce the traveling wave ansatz, namely
\begin{equation}
    r(X,T) = R\left(Z\right),\qquad Z = X - cT.
    \label{Traveling wave ansatz}
\end{equation}
Substituting this ansatz in \eqref{Target PDE model}, integrating twice 
(using a suitable integrating factor)
and setting an arbitrary integration constant to zero as appropriate when looking for bounded solutions, yields
\begin{equation}
    \frac{\epsilon^{2}c^{2}}{12}\left(R'\right)^{2} = -\frac{2}{p+1}R^{p+1} + c^{2}R^{2} - 2aR - b,
    \label{final traveling wave ODE}
\end{equation}
where $'$ denotes differentiation with respect to $Z$ throughout this section,
and $a, b$ are two integration constants.
When $p$ takes on half-integer values, 
it is convenient to apply the transformation $R = g^{2}$ so that Eq.~\eqref{final traveling wave ODE} 
becomes
\begin{equation}
    \frac{\epsilon^{2}c^{2}}{3}g^{2}\left(g'\right)^{2} = -\frac{2}{p+1}g^{2\left(p+1\right)} + c^{2}g^{4} - 2ag^{2} -b.
    \label{transformed ODE}
\end{equation}
To compute the associated traveling wave solution, we need $b = 0$, given its decay at
infinity. Then,  we can divide both sides of Eq~\eqref{transformed ODE} by $g^{2}$ to obtain 
\begin{equation}
    \frac{\epsilon^{2}c^{2}}{3}\left(g'\right)^{2} = -\frac{2}{p+1}g^{2p} + c^{2}g^{2} - 2a.
    \label{transformed ODE 2}
\end{equation}
To compute the associated traveling wave solution to 
Eq.~\eqref{transformed ODE 2}, the decay of the
solution again requires $a = 0$ so that the equation becomes separable and the solution,
obtained via quadrature, reads
\begin{equation}
    g(Z) = \left[\frac{(p+1)c^{2}}{2}\right]^{\frac{1}{2(p-1)}}\text{sech}^{\frac{1}{p-1}}\left(\frac{\sqrt{3}(p-1)}{\epsilon}(Z-Z_{0})\right),
    \label{traveling wave solution for g}
\end{equation}
where $Z_0$ is an integration constant.
Indeed, the relevant calculation arises in the
process of identifying the bright solitary
waves of the general power variant of the
nonlinear Schr{\"o}dinger equation~\cite{sulem}.
Recall that since $R = g^{2}$, the corresponding traveling wave solution for $R$ assumes the form
\begin{equation}
    R(Z) = \left[\frac{(p+1)c^{2}}{2}\right]^{\frac{1}{p-1}}\text{sech}^{\frac{2}{p-1}}\left(\frac{\sqrt{3}(p-1)}{\epsilon}(Z-Z_{0})\right).
    \label{traveling wave solution for u}
\end{equation}
Written in terms of the original granular (strain) variable, this becomes,
\begin{equation}
    r_n(t) = \left[\frac{\left(p+1\right)c^{2}}{2}\right]^{\frac{1}{p-1}}\text{sech}^{\frac{2}{p-1}}\left(\sqrt{3}\left(p-1\right)\left(n - ct\right)\right)
    \label{traveling wave solution for r}.
\end{equation}
Note that the solitary wave approximation is independent of the parameter
$\epsilon$.

\subsection{Comparison of traveling waves of different models}

We now perform a comparison of the traveling wave solutions corresponding to 
different approximate models for the granular chain, including Eq.~\eqref{Target PDE model}, the 
model of Eq.~(\ref{not good model})
proposed in the work of~\cite{Ahnert_2009}
and finally the earliest continuum approximation
from the classic work of~\cite{Nester2001}
(through the continuum limit in displacements).
The work of~\cite{Ahnert_2009} already compared
the latter two continuum models with the
numerically exact traveling lattice solution.
For completeness, we also remind
the reader the form of the traveling wave
solution for these
other 3 models (2 continuum approximations, namely \cite{Ahnert_2009} and 
\cite{Nester2001}, as well as for the original 
lattice model) in what follows.

Using the traveling wave ansatz \eqref{Traveling wave ansatz} in Eq.~\eqref{not good model} of~\cite{Ahnert_2009}, we reduce it to the following ODE,
\begin{equation}
    c^{2}R = R^{p} + \frac{\epsilon^{2}}{12}\left(R^{p}\right)''.
    \label{Pikovsky's regularized model}
\end{equation}
The solution to Eq.~\eqref{Pikovsky's regularized model} reads
\begin{equation}
    R(Z) = |c|^{m}A_{1}\cos^{m}\left(B_{1}Z\right),
    \label{Pikovsky's TW}
\end{equation}
where
\begin{equation}
    m = \frac{2}{p-1},\qquad 
    A_{1} = \left(\frac{p+1}{2p}\right)^{\frac{1}{1-p}}, \qquad
    B_{1} = \frac{\sqrt{3}}{\epsilon}\frac{p-1}{p}.
\end{equation}
Importantly, note that Eq.~\eqref{Pikovsky's TW} applies only between two consecutive zeros of the cosine, 
beyond which $R(z)$ is taken to be identically zero.  The same is true for Eq.~\eqref{Nesterenko's traveling wave} below.
Similarly, the classical continuum PDE model of~\cite{Nester2001}
reads
\begin{equation}
    r_{TT} = \left(r^{p}\right)_{XX} + \frac{\epsilon^{2}}{12}\left(\left(r^{p}\right)_{XXXX} - \frac{p(p-1)}{2}\left(r^{p-2}r^{2}_{X}\right)_{XX}\right).
    \label{Pikovsky's continuum model}
\end{equation}
Substituting the traveling wave ansatz \eqref{Traveling wave ansatz} into Eq.~\eqref{Pikovsky's continuum model} yields
\begin{equation}
    c^{2}R = R^{p} + \frac{\epsilon^{2}}{12}\left(R^{p}\right)'' - \frac{\epsilon^{2}p(p-1)}{24}R^{p-2}\left(R'\right)^{2}.
    \label{Nesterenko's traveling wave ODE}
\end{equation}
The solution to Eq.~\eqref{Nesterenko's traveling wave ODE} reads
\begin{equation}
    R(Z) = |c|^{m}A_{2}\cos^{m}\left(B_{2}Z\right),
    \label{Nesterenko's traveling wave}
\end{equation}
where
\begin{equation}
    m = \frac{2}{p-1}, 
    \qquad 
    A_{2} = \left(\frac{2}{1+p}\right)^{\frac{1}{1-p}}, 
    \qquad 
    B_{2} = \sqrt{\frac{6}{\epsilon^{2}}\frac{(p-1)^{2}}{p(p+1)}}.
\end{equation}

As discussed above,
we also wish to compare the results of
the proposed model~\eqref{Target PDE model} 
with the numerically computed exact traveling wave solution of the original granular chain. These are solutions
of the form $r_n(t) = r(n- ct) = r(z)$ where $r(z)$ satisfies the advance-delay equation
\begin{equation} \label{advdelay}
    c^2 r'' = r^p(z-1) - 2 r^p(z) + r^p(z+1).
\end{equation}
To this end, we applied an iterative algorithm \cite{HOCHSTRASSER1989259} to numerically compute the solutions of Eq.~\eqref{advdelay}, see also
\cite{granularBook}.  Note that when we refer to the ``exact"  solution, we are
referring to the numerical solution found to a specified numerical tolerance
of Eq.~\eqref{advdelay}.

\begin{figure}[ht]
    \centering
    \includegraphics[width=1.1\linewidth]{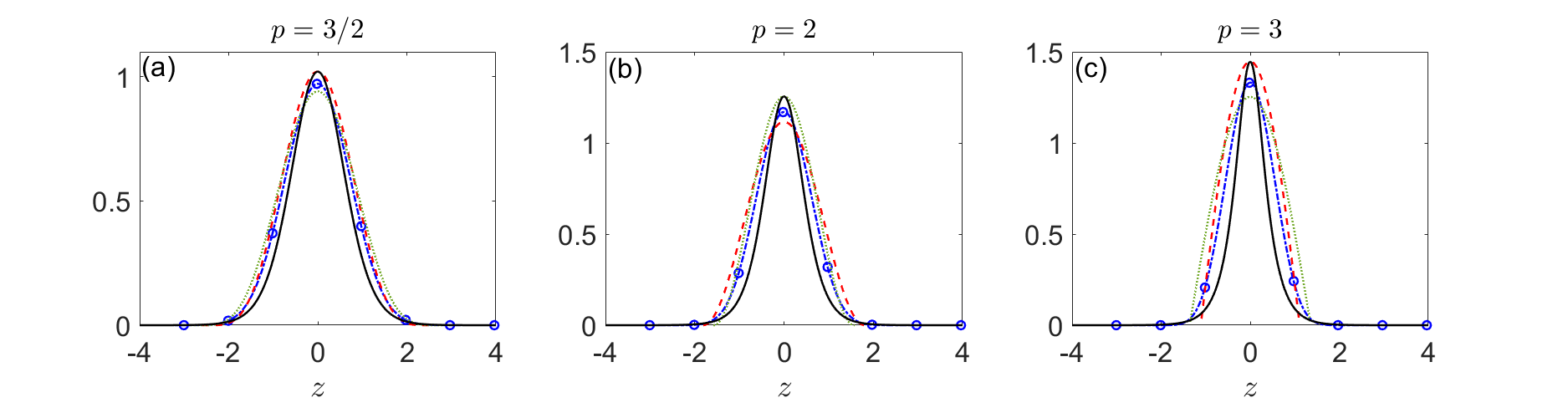}
    \includegraphics[width=1.1\linewidth]{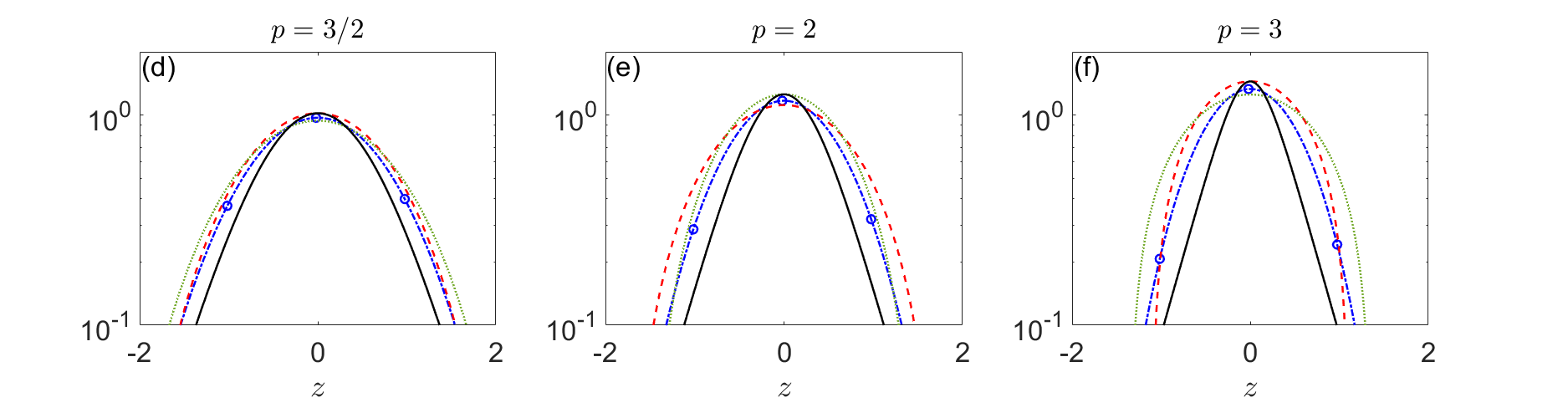}
    \caption{Comparison of the traveling solitary waves in different continuum models for different values of the parameter $p$: The leftmost, middle, and rightmost columns denote the cases of $p = 3/2, 2, 3$, respectively. The blue lines (with circles) denote the ``exact" solitary of Eq.~\eqref{advdelay}, and the red dashed line, green dotted line, black solid line refer to the solitary waves associated with the Ahnert-Pikovsky \eqref{not good model}, Nesterenko \eqref{Pikovsky's continuum model}, and the regularized continuum model \eqref{Target PDE model}, respectively. The first row shows the comparison of the solitary waves of different continuum models in their respective standard scale, while the second row depicts the semi-log scale of all solitary waves. Note that Ahnert-Pikovsky's and Nesterenko's approximations for the solitary waves are only plotted over the period of the respective cosine.}
    \label{fig: Solitary Wave Comparison}
\end{figure}

Figure \ref{fig: Solitary Wave Comparison} showcases the numerical comparison of 4 traveling wave solutions (3 approximations and 1 ``exact" ---again, to
a prescribed tolerance---) for the cases of $p = 3/2$, $2$, $3$.
 
Each of the traveling wave solutions which arise in periodic (cosinusoidal) form is plotted only for one period {of the corresponding cosine function, 
beyond which the solution is taken to be zero.} 
Note that, while the continuum models themselves depend on the parameter
$\epsilon$, the corresponding predictions is independent of $\epsilon$ once returning
to the original lattice variables, see e.g. Eq.~\eqref{traveling wave solution for r}. 
Thus, we do not need to ``select" any particular value of $\epsilon$ when 
comparing solitary waves.

For the values of $p$ considered here, the approximation from~\eqref{traveling wave solution for r} 
is meaningfully proximal to the exact solution,
as are the other models. 
See the black solid curve of Fig.~\ref{fig: Solitary Wave Comparison}. In terms of the error $E=\frac{1}{N}\sum_n^N |r_n - R(\epsilon n)|$, where $r_n$ is an ``exact" traveling wave
and $R$ a traveling wave from one of the PDE models,
the approximation errors
are of the same magnitude. For example, with $p=3/2$ the errors are
$E = 0.034$ for the strain variant of the model of Nesterenko \eqref{Pikovsky's continuum model}, $E = 0.030$ for the model of Ahnert-Pikovsky \eqref{not good model},
and $E = 0.046$ for the model of considered in this paper, namely the regularized model of \eqref{Target PDE model}. The errors
for other values of $p$ are similar. We notice that although the error of the model \eqref{Target PDE model} is the greatest, it is still quite small and also comparable to the other two. Recall that the solitary
wave solution of the granular chain has a double exponential decay in the tails \cite{Chatterjee,pego1},
whereas the approximation from \eqref{traveling wave solution for r} has only
exponential decay. The approximations  \eqref{Pikovsky's TW} and \eqref{Nesterenko's traveling wave}
have finite support. See Fig.~\ref{fig: Solitary Wave Comparison}(d-f), which shows the solutions
in a semi-log scale where the decay rate can be better discerned.  While the quantitative differences
between the actual solution and approximation are similar for all cases,
it still remains an interesting open question if, in terms of rigorous error bounds,
any of the three considered here is the ``most" accurate. It is worth mentioning that here we
have only examined the stationary (in the co-traveling
frame) aspect of solitary waves. Towards the
end of our presentation, we will return to the
dynamical properties of these models, as concerns
the prototypical structure considered herein,
namely the DSW.

\section{Periodic traveling wave solutions}\label{s:periodicwaves}

Next, we look for periodic solutions, which as usual will play a crucial role in the study of DSWs. 
In particular, we seek solutions to Eq.~\eqref{transformed ODE 2} with $a \neq 0$. 
Unlike the solitary wave solutions, however, where we were able to obtain solutions for arbitrary values of~$p$, 
in this case the calculations (and the resulting expressions) are heavily dependent on the specific value of~$p$.
Here we will focus on three special cases: $p = 3/2$, $p = 2$, and $p = 3$. In the following three subsetions, we only list the analytical expression of the periodic wave solutions for each of the three cases and defer the detailed derivation of these solutions to Appendix~A.

\subsection{Case $p = 3/2$}

When $p = 3/2$, the original ODE \eqref{transformed ODE 2} becomes
\begin{equation}
    \frac{\epsilon^{2}c^{2}}{3}\left(g'\right)^{2} = -\frac{4}{5}g^{3} + c^{2}g^{2} - 2a,
    \label{transformed ODE 2 for p = 3/2}
\end{equation}
where, as before, primes denote differentiation with respect to $Z$.

A direct integration of Eq.~\eqref{transformed ODE 2 for p = 3/2} yields,
\begin{equation}\label{eq: periodic sol to p = 3/2 at g level}
    g(z) = g_2 + \left(g_3 -g_2\right)\text{cn}^{2}\left( \frac{\sqrt{3\left(g_3 - g_1\right)}}{\sqrt{5}\epsilon c}\left(Z-z_0\right), m\right)
\end{equation}
where $\text{cn}\left(Z,m\right)$ denotes the Jacobi elliptic cosine function with parameter $m$ ($\sqrt{m}$ is the modulus), and $g_1, g_2, g_3$ the three roots of the potential curve $P\left(g\right) = -g^{3} + \frac{5}{4}c^{2}g^{2} - \frac{5}{2}a$, and $m = \frac{g_3 - g_2}{g_3 - g_1}$.

Finally, since $R = g^{2}$, we solve for $R$ to get the following periodic solution,
\begin{equation}\label{eq: periodic solution for p = 3/2 -- main body}
    R(Z) = \left[g_{2} + \left(g_{3} - g_{2}\right)\text{cn}^{2}\left( \frac{\sqrt{3\left(g_3 - g_1\right)}}{\sqrt{5}\epsilon c}\left(Z-z_0\right), m\right)\right]^{2}.
\end{equation}

Note that this is only a two-parameter family of periodic waves, and hence is not the most general form. In the derivation of this formula,
one of the integration constants was set to zero, see
Eq.~\eqref{transformed ODE 2 for p = 3/2}. We were unable
to find a analytical formula in the general, three parameter, case.

\subsection{Case $p = 2$}

When $p = 2$ we do not need to apply the transformation $R = g^2$, so we simply focus on the original ODE \eqref{final traveling wave ODE} which now becomes
\begin{equation}
    \frac{\epsilon^{2}c^{2}}{12}\left(R'\right)^{2} = c^{2}R^{2} - \frac{2}{3}R^{3} - 2BR - C,
    \label{ODE for p = 2 4.4}
\end{equation}
where we renamed the two constants of integration $\left(a,b\right) = \left(B,C\right)$.

Note, in contrast to the previous case, this ODE has three free parameter $(B,C,c)$. A direct integration of Eq.~\eqref{ODE for p = 2 4.4} yields
\begin{equation}
    R(Z) = R_{2} + \left(R_{3}-R_{2}\right)\text{cn}^{2}\left( \frac{\sqrt{2\left(R_{3}-R_{1}\right)}}{\epsilon c}\left(Z-z_{0}\right), m\right),
    \label{periodic solution for the case p =2 -- main body}
\end{equation}
where $R_1 < R_2 < R_3$ denote the three roots of the potential $P(R) = -R^{3}+\frac{3}{2}c^{2}R^{2}-3BR-\frac{3}{2}C$, and
\begin{equation}
m = \frac{R_3 - R_2}{R_3 - R_1}.
\end{equation}
Moreover, we can also deduce the leading-edge soliton amplitude through the periodic solution expression in \eqref{periodic solution for the case p =2}. Namely, at the solitonic limit $m \to 1$, so $R_2 \to R_1$ and the periodic solution reduces to the hyperbolic secant function with background given by $R_2$ and amplitude denoted by $a^{+}$,
\begin{equation}\label{eq: soliton amplitude for p = 2}
    a^{+} = R_3 - R_2.
\end{equation}
It will be convenient  to express the soliton amplitude in terms of the wave's background, which we call $r^+$, instead of the two unknown parameters $R_3$ and $R_2$. 
As we will see later in Sec.~\ref{s:riemann}, $r^+$ represents the right value of the strain in the Riemann problem involving a jump from $r^{-}$ to $r^{+}$ (see also Eq.~\eqref{step}).
Since at the soliton limit $R_2 = R_1$ and since $R_2 = r^{+}$ is the background, we have that $R_1=R_2 = r^{+}$. On the other hand, to determine the unknown parameter $R_3$, we expand the polynomial product $\left(R_1-R\right)\left(R_2-R\right)\left(R_3 - R\right)$ and then equate coefficients with the polynomial of $P\left(R\right)$ to obtain that
\begin{equation}\label{eq: equating coefficients}
    R_1 + R_2 + R_3 = \frac{3}{2}c^{2}.
\end{equation}
where $c$ now denotes the theoretically predicted 
speed. Given the information that $R_1 = R_2 = r^{+}$, we solve for $R_3$ in terms of $R_1, R_2$ and $c$ and finally substitute it into Eq.~\eqref{eq: soliton amplitude for p = 2} to arrive at the following explicit formula for the 
soliton amplitude,
\begin{equation}\label{eq: explicit soliton amplitude formula for the case p = 2}
    a^{+} = \frac{3}{2}c^{2} - 3r^{+}.
\end{equation}

\subsection{Case $p = 3$}

For the case $p = 3$, we first rewrite the traveling ODE as follows,
\begin{equation}
    \left(R'\right)^{2} = -\frac{6}{\epsilon^{2}c^{2}}\left(R^{4}-2c^{2}R^{2}+4aR+2b\right)
    = -\frac{6}{\epsilon^{2}c^{2}}\left(R-R_1\right)\left(R-R_2\right)\left(R-R_3\right)\left(R-R_4\right).
    \label{Rewriting the traveling ODE for p = 3 version 2}
\end{equation}
We denote
\begin{equation}
    \mu = -\frac{6}{\epsilon^{2}c^{2}}.\label{Definition of the diffusion term}
\end{equation}
Clearly $\mu < 0$, and then we first make the assumption that all four roots of $R_1, R_2, R_3, R_4$ are real valued and further assume the following order of the four roots,
\begin{equation}
    R_1 \leq R_2 \leq R_3 \leq R_4.
    \label{order of the 4 roots}
\end{equation}
and also assume that the oscillation occurs in the interval $R_3 \leq R \leq R_4$.
Integration of Eq.~\eqref{Rewriting the traveling ODE for p = 3 version 2} then yields
\begin{equation}
    R = R_3 + \frac{\left(R_4 - R_3\right)\text{cn}^{2}\left(\zeta, m\right)}{1 + \frac{R_4 - R_3}{R_3 - R_1}\text{sn}^{2}\left(\zeta, m\right)},
    \label{Periodic solution for the case p = 3 -- main body}
\end{equation}
where
\bse
\begin{align}
    m &= \frac{\left(R_4 - R_3\right)\left(R_2 - R_1\right)}{\left(R_4 - R_2\right)\left(R_3 - R_1\right)},
    \label{Elliptic modulus m_2}\\
    \zeta &= \frac12{\sqrt{\left|\mu\right|\left(R_3 - R_1\right)\left(R_4 - R_2\right)}Z}. 
    \label{Phase variable}
\end{align}
\ese
In the soliton limit where $R_3 \to R_2$, again by \cite{El_2017}, we obtain the following soliton solution:
\begin{equation}
    R = R_2 + \frac{R_4 - R_2}{\text{cosh}^{2}\zeta + \frac{R_4 - R_2}{R_2 - R_1}\text{sinh}^{2}\zeta},
    \label{Soliton solution for p = 3}
\end{equation}
So we know immediately from the soliton solution \eqref{Soliton solution for p = 3} that the soliton amplitude reads
\begin{equation}
    a^+ = R_4 - R_2.
    \label{Soliton amplitude for p = 3}
\end{equation}
which is completely analogous to the case of $p = 2$.

Finally, to get an explicit analytical formula for the soliton amplitude, we notice that by expanding the product of polynomials of \eqref{Rewriting the traveling ODE for p = 3 version 2} and equating the relevant coefficients, we have that
\bse
\begin{align}
    R_1 + R_2 + R_3 + R_4 &= 0,
    \label{result of equating coefficient 1}\\
    R_1R_2 + R_1R_3 + R_2R_3 + R_1R_4 + R_2R_4 + R_3R_4 &= -2c^{2}.
    \label{result of equating coefficient 2}
\end{align}
\ese
Because we are at the soliton limit, we substitute the relation $R_3 = R_2$ into the system of \eqref{result of equating coefficient 1}-\eqref{result of equating coefficient 2} to obtain that
\begin{align}
    R_1 + 2R_2 + R_4 &= 0,\label{result of equating coefficient 1 after substitution}\\
    2R_1R_2 + R_2^{2} +R_1R_4 + 2R_2R_4 &= -2c^{2}.
    \label{result of equating coefficient 2 after substitution}
\end{align}
We then eliminate $R_1$ from the system of \eqref{result of equating coefficient 1 after substitution}-\eqref{result of equating coefficient 2 after substitution} to have that
\begin{equation}
    R_4^{2} + 2R_4R_2 + \left(3R_2^{2} - 2c^{2}\right) = 0.
    \label{ultimate quadratic equation}
\end{equation}
The background is once again $R_2=r^{+}$.
Then we solve \eqref{ultimate quadratic equation} for $R_4$ to obtain that
\begin{equation}
    R_4^{\pm} = -R_2 \pm \sqrt{2\left(c^{2}-R_2^{2}\right)}.
    \label{solution for u_4}
\end{equation}
Here, we need to take the root $R_4^{+}$ and ignore $R_4^{-}$ to avoid the issue of negative soliton amplitude.

Finally, substituting $R_4^{+}$ into Eq.~\eqref{Soliton amplitude for p = 3} we obtain an explicit soliton amplitude formula for $p = 3$,
\begin{equation}
    a^{+} = \sqrt{2\left(c^{2} - \left(r^{+}\right)^{2}\right)} - 2r^{+}.
    \label{Explicit formula for soliton amplitude for p = 3}
\end{equation}

\section{Conservation laws}\label{s:conservationlaws}

Recall that the continuum model \eqref{Target PDE model} is an approximation of the discrete granular chain at the level of the strain. Interestingly, this continuum model can be transformed into its associated displacement version which is an approximation model for the discrete system \eqref{Displacement ODE}. Denoting the displacement variable for the PDE $u(X,T)$, the relationship between the displacement $u(X,T)$ and the strain $r(X,T)$ is 
\begin{equation}
    r(X,T) = u_X(X,T).
    \label{strain-displacement relation}
\end{equation}
which then in turn would approximate the displacement of the granular chain via
$u_n(t) \approx -\epsilon^{-1} u(X,T)$. Note that we need to include the negative
sign since the strain variable is $r_n = u_{n-1}-u_{n}$, which has the
opposite sign of the difference as compared to the one associated to the spatial derivative.

To obtain the continuum model in terms of $u(X,T)$, we substitute \eqref{strain-displacement relation} into the original continuum model \eqref{Target PDE model} and then observe that
\begin{equation}
    u_{XTT} - \frac{\epsilon^{2}}{12}u_{XXXTT} = \left[\left(u_X\right)^{p}\right]_{XX}.
    \label{Substitution of u}
\end{equation}
Integrating Eq.~\eqref{Substitution of u} with respect to $X$ (assuming the integration constant to be zero)
gives
\begin{equation}
    u_{TT} - \frac{\epsilon^{2}}{12}u_{XXTT} = \left[\left(u_X\right)^{p}\right]_{X}.
    \label{continuum model at displacement}
\end{equation}
Interestingly, the displacement continuum model \eqref{continuum model at displacement} has a few conservation laws, namely the conservation of momentum and the conservation of energy. These two conservation laws can be seen by the following two rearrangements of Eq.~\eqref{continuum model at displacement},
\begin{equation}
    \left(u_{X}u_{T}+\frac{\epsilon^{2}}{12}u_{XT}u_{XX}\right)_{T} - \left(\frac{1}{2}\left(u_{T}\right)^{2}+\frac{\epsilon^{2}}{24}\left(u_{XT}\right)^{2}+\frac{p}{p+1}\left(u_{X}\right)^{p+1}+\frac{\epsilon^{2}}{12}u_{XTT}u_{X}\right)_{X} = 0.
    \label{Conservation of momentum}
\end{equation}
and 
\begin{equation}
    \left(\frac{1}{2}\left(u_{T}\right)^{2}+\frac{\epsilon^{2}}{24}\left(u_{XT}\right)^{2}+\frac{1}{p+1}\left(u_{X}\right)^{p+1}\right)_{T} - \left(\frac{\epsilon^{2}}{12}u_{XTT}u_{T}+\left(u_{X}\right)^{p}u_{T}\right)_{X} = 0,
    \label{Conservation of energy}
\end{equation}
respectively. We notice that Eq.~\eqref{Conservation of momentum} corresponds to the conservation of linear momentum, while Eq.~\eqref{Conservation of energy} refers to the conservation of energy. In addition, it is also worthwhile to note that for the particular case when $\epsilon = 0$ and $p = 1$,  Eq.~\eqref{continuum model at displacement} simply reduces to the familiar linear wave equation where this situation immediately falls back to a standard exercise regarding the two conservation laws (see, e.g., \cite{strauss2007partial}).

At the strain level, the PDE \eqref{Target PDE model} yields an equivalent conservation law
to \eqref{continuum model at displacement}:
\begin{equation}
    \left(r_{T} - \frac{\epsilon^{2}}{12}r_{XXT}\right)_{T} = \left[\left(r^{p}\right)_{X}\right]_{X}.
    \label{conservation of the PDE itself}
\end{equation}
The above conservation laws can be used to derive the Whitham modulation equations (e.g., see \cite{Mark2016}), 
but in the following section we will use an alternative approach 
based on Lagrangian formulation of the PDE~\eqref{Target PDE model}.

\section{Whitham modulation equations}\label{s:whitham}

Ever since the seminal work of Whitham~\cite{PRSA283p238,Whitham74} and of Gurevich and Pitaevskii \cite{JETP38p291}, 
the method of Whitham modulation theory 
has proved to be an effective tool for characterizing DSWs quantitatively   
(e.g., see \cite{Mark2016} for a review of this subject).
The main object of study in Whitham modulation theory is to derive the so-called Whitham modulation equations, which govern spatio-temporal modulations of the periodic solutions 
of the model in question.
In this section we derive the Whitham modulation equations for the PDE~\eqref{Target PDE model} that approximates the granular chain in the continuum limit. Here we only consider the case of $p = 2$ since the derivation of the modulation equations for other cases of $p$ is completely analogous, see Appendix B.

\subsection{Theoretical preliminaries}

The idea of modulation theory is to consider slow modulations of the parameters that 
completely determine the periodic solutions and derive their associated governing equations. 
To this end, we first introduce the wavenumber $K$ and frequency $\Omega$ to rewrite the traveling-wave ansatz as $r(X,T) = R(\theta)$, 
with the phase variable $\theta = (KX-\Omega T)/\epsilon$ where $\Omega = c_p K$,
to express our periodic solutions.

In the case of $p=2$, the periodic
solution in terms of $\theta$ is,
\begin{equation}
    R\left(\theta\right) = R_{2} + \left(R_{3}-R_{2}\right)\text{cn}^{2}\left( \frac{\sqrt{2\left(R_{3}-R_{1}\right)}}{c_pK}\left(\theta - \theta_0 \right), m\right),
    \label{scaled periodic solution}
\end{equation}
where $\theta_0$ is an arbitrary constant of integration which will be set to zero in the following.
Note, Eq.~\eqref{scaled periodic solution} is equivalent to the one shown in Eq.~\eqref{ODE for p = 2 4.4} when returning to the $r(X,T)$ variables
(in particular, the roots $R_1,R_2,R_3$
are the same despite the change
of variable from $Z$ to $\theta$).

We now rewrite the periodic solution \eqref{scaled periodic solution} so that its period is fixed, i.e., is
independent of the solution parameters,
which is needed in the derivation of the modulation equations.
To this end, we use the fact that the periodic solution oscillates between the two values $R_2 < R_3$, and observe that
\begin{equation}
    2\pi = \int_{0}^{2\pi} \text{d}\theta = 2\int_{R_2}^{R_3}\frac{dR}{R_{\theta}} 
    = 2\int_{R_2}^{R_3}\frac{dR}{\sqrt{\frac{8}{K^{2}{c_p}^{2}}\left(R_1 - R\right)\left(R_2 - R\right)\left(R_3 - R\right)}} = \frac{2 c_pKK_m}{\sqrt{2\left(R_3 - R_1\right)}},
\end{equation}
where $K_m$ is the complete elliptic integral of the first kind. 
Thus, we have
\begin{equation}
    \frac{K_m}{\pi} = \frac{\sqrt{2\left(R_3 - R_1\right)}}{c_pK}.
    \label{derived fact}
\end{equation}
Using the independent variable $\theta$ as defined above, we then obtain that the periodic solution \eqref{scaled periodic solution} is a $2\pi$-periodic function
\begin{equation}
    R\left(\theta\right) = R_{2} + \left(R_{3}-R_{2}\right)\text{cn}^{2}\left(\frac{K_m}{\pi}\theta, m\right).
    \label{reparametrized periodic sol}
\end{equation}

Furthermore, it is also convenient to reparametrize the solution, which is done by expressing the parameters $R_1,R_2, R_3$ in terms of $K,m,c_p$ using the following relations
\bse
\label{e:R1R2R3m}
\begin{align}
    &R_3 - R_1 = \frac{K^{2}c_p^{2}K_m^{2}}{2\pi^{2}}, 
    \label{first relation eqn}
    \\
    &R_1 + R_2 + R_3 = \frac{3c_p^{2}}{2}, 
    \label{second relation eqn}
    \\
    &m = \frac{R_3 - R_2}{R_3 - R_1},
    \label{third relation eqn}
\end{align}
\ese
where \eqref{first relation eqn} comes from \eqref{derived fact} and \eqref{second relation eqn} from the relation
\begin{equation}
    (R_1 - R)(R_2 - R)(R_3 - R) = -R^{3} + \frac{3}{2}c_p^{2}R^{2} - 3BR -\frac{3}{2}C.
    \label{equation for the polynomial}
\end{equation}
Then, solving the system \eqref{e:R1R2R3m} for $R_1, R_2, R_3$ in terms of $K,c_p,m$ yields
\bse
\label{e:R1R2R3new}
\begin{align}
    R_1 &= \frac{c_p^{2}}{2} + \frac{\left(m-2\right)K^{2}c_p^{2}K_m^{2}}{6\pi^{2}},
    \label{solution for u_1}
    \\
    R_2 &= \frac{c_p^{2}}{2} + \frac{\left(1-2m\right)K^{2}c_p^{2}K_m^{2}}{6\pi^{2}},
    \label{solution for u_2}
    \\
    R_3 &= \frac{c_p^{2}}{2} + \frac{\left(m+1\right)K^{2}c_p^{2}K_m^{2}}{6\pi^{2}}.
    \label{solution for u_3}
\end{align}
\ese
Substituting \eqref{e:R1R2R3new} into the reparametrized periodic solution \eqref{reparametrized periodic sol} yields
\begin{equation}
    R\left(\theta\right) = \frac{c_p^{2}}{2} + \frac{\left(1-2m\right)K^{2}c_p^{2}K_m^{2}}{6\pi^{2}} + \frac{mK^{2}c_p^{2}K_m^{2}}{2\pi^{2}}\text{cn}^{2}\left(\frac{K_m}{\pi}\theta, m\right),
    \label{ultimate reparametrized periodic sol}
\end{equation}
where now clearly the periodic solution is parametrized by the three parameters $c_p, K, m$.

\subsection{Derivation of modulation equations}
 
To derive the modulation equations, we will use the method of averaged Lagrangian~\cite{Kamchatnov}.  
We first note that the PDE model in its displacement form, see Eq.~\eqref{continuum model at displacement}, can be obtained through a variational principle. We observe that the Lagrangian density $\mathbb{L}$ associated with Eq.~\eqref{continuum model at displacement} is

\begin{equation}
\label{eq: Lagrangian}
    \mathbb{L} = \frac{1}{2}\left(u_T\right)^{2} + \frac{\epsilon^{2}}{24}u_{XX}u_{TT} - \frac{\left(u_X\right)^{p+1}}{p+1}.
\end{equation}

Equation~\eqref{continuum model at displacement} represents the Euler-Lagrange equation for the action functional 
$\iint\mathbb{L} \, dX dT$.
It is worth noting that the middle
term could be replaced by
$(\epsilon^2/24)u^2_{XT}$, which could
be interpreted as a “microkinetic energy”~\cite{theil}, although we 
will not pursue that hereafter.

A modulated traveling wave is an approximate solution whose parameters vary slowly relative to a fast phase
$\theta(X,T)$ and a so-called fast pseudo phase $Q(X,T)$. The ansatz is formulated at the level of the displacement
and has the form
 \begin{equation} \label{Variational ansatz}
     u(X,T) = \epsilon\left( Q(X,T) + \psi\left( \theta(X,T) \right) \right) \hspace{5mm} 
 \end{equation}
 where,
\bse \label{fast phases}
\begin{align}
    \theta_X = K(X,T)/\epsilon\,,\qquad \theta_T = - \Omega(X,T)/\epsilon\,,
    \label{real phase theta}\\
    Q_X = \beta(X,T)/\epsilon\,,\qquad Q_T = - \gamma(X,T)/\epsilon\,,
    \label{pseudo phase eta}
\end{align}
\ese
and where $\psi(\theta)$ a $2\pi$-periodic function with zero average,
namely $\overline{\psi} = 0$, 
where the bar denotes the averaging operation over a period of the function,
\begin{equation}\label{eq: averaging operation}
    \overline{f} = \frac{1}{2\pi}\int_0^{2\pi}f\left(\theta\right)\,d\theta
.
\end{equation}
Note that, with this ansatz, all terms can be expressed
in terms of the traveling wave at the strain level $R$.
In particular, $R=R(\theta)$ satisfies
\begin{equation}
    K^{2}R_\theta^{2} = -\frac{24}{\left(p+1\right)c_p^{2}}R^{p+1} + 12R^{2} - 24BR - 12C,
    \label{PDE at the co-traveling frame}
\end{equation}
where $c_p = \Omega/K$ and $B, C$ are two constants of integration. Equation~\eqref{PDE at the co-traveling frame} is
analogous to Eq.~\eqref{ODE for p = 2 4.4}, but the $\epsilon$
has vanished due to the scaling of $\theta$. 
Note that $R(\theta)$ relates to $u(X,T)$
and its derivatives in the following way,
\bse
\label{u and R}
\begin{align} 
u_X &= \beta + \psi'(\theta) K = R(\theta)\\
u_T &= -\gamma - \psi'(\theta)\Omega = -\gamma - (R(\theta) - \beta)c_p\\
u_{XX} &= \psi''(\theta)K^2/\epsilon =  R_\theta(\theta)K/\epsilon \\
u_{TT} &= \psi''(\theta)\Omega^2/\epsilon =  R_\theta(\theta) \Omega^2/( K \epsilon) 
\end{align}
\ese
where we have used Eq.~\eqref{fast phases}.
Now we are ready to derive modulation equations.
Substituting \eqref{Variational ansatz} into the Lagrangian density of Eq.~\eqref{eq: Lagrangian} and expressing everything in terms of $R$
using Eqs.~\eqref{PDE at the co-traveling frame} and \eqref{u and R} yields
\begin{equation}
    \mathbb{L} = \frac{\Omega^{2}}{12}\left(R_\theta\right)^{2} +\left(-\beta c_p^{2}+\gamma c_p + Bc_p^{2}\right)R + \frac{1}{2}\gamma^{2}+\frac{1}{2}c_p^{2}\beta^{2}-\beta\gamma c_p + \frac{1}{2}Cc_p^{2}.
    \label{explicit Lagrangian}
\end{equation}
The method of the averaged Lagrangian assumes that the
wave parameters are constant over one period of motion.
We therefore compute the average Lagrangian,

\begin{equation}
    \mathcal{L} = \frac{1}{2\pi}\int_{0}^{2\pi}\mathbb{L}\,d\theta 
    = \frac{\Omega c_p}{12}W\left(B,C,c_p\right)+\frac{1}{2}\gamma^{2}-\frac{1}{2}c_p^{2}\beta^{2}+\frac{1}{2}Cc_p^{2}+\beta Bc_p^{2},
    \label{average Lagrangian}
\end{equation}
where we have the "action" integral $W\left(B,C,c_p\right)$ defined as follows

\begin{equation}
    W\left(B,C,c_p\right) = \frac{K}{2\pi} \int_{0}^{2\pi}\left(R_\theta\right)^{2}d\theta 
    = \frac{1}{2\pi}\oint\left(-\frac{24}{\left(p+1\right)c_p^{2}}R^{p+1}+12R^{2}-24BR - 12C\right)^{1/2}dR.
    \label{action integral}
\end{equation}

The modulation system then simply follows from the average variational principle,
\begin{equation}
    \delta\iint\mathcal{L}\left(K,\Omega, \beta, \gamma, B, C\right)\,dXdT = 0.
    \label{variational principle}
\end{equation}
which then yields the following Euler-Lagrange equations
(and corresponding consistency relations),
\bse
\begin{align}
     &\mathcal{L}_B = 0, \hspace{5mm}\mathcal{L}_C = 0,\label{constant variations}\\
     &\frac{\partial}{\partial T}\mathcal{L}_\Omega -\frac{\partial}{\partial X}\mathcal{L}_K = 0, \hspace{5mm} K_{T} + \Omega_{X} = 0,\label{real phase variations}\\
    &\frac{\partial}{\partial T}\mathcal{L}_\gamma - \frac{\partial}{\partial X}\mathcal{L}_\beta = 0, \hspace{5mm} \beta_{T} + \gamma_{X} = 0\,.
    \label{real phase variation 2}
\end{align}
\ese
Using equations \eqref{average Lagrangian}, we obtain from \eqref{constant variations} that
\bse
\begin{align}
    &\beta = -\frac{\Omega}{12c_p}W_B \label{explicit constant variations 1}\\
    &\frac{\Omega c_p}{12}W_C + \frac{1}{2}c_p^{2} = 0.
    \label{explicit constant variations 2}
\end{align}
\ese
Equation \eqref{explicit constant variations 1} simplifies to 
\begin{equation}
    \beta = \overline{R},
    \label{recover beta from variational principle}
\end{equation}
which simply states the average of the strain profile
of the wave is $\beta$, a fact that is obvious
by the construction of the ansatz Eq.~\eqref{Variational ansatz}.
For Eq.~\eqref{explicit constant variations 2}, if we solve for the wavenumber $K$, we end up with
\begin{equation}
    K = -\frac{6}{W_C}.
    \label{solution for wavenumber}
\end{equation}
Then, the nonlinear dispersion relation reads
\begin{equation}
    \Omega = Kc_p = -\frac{6c_p}{W_C}.
    \label{nonlinear dispersion relation}
\end{equation}

{We observe that the two equations of \eqref{recover beta from variational principle} and \eqref{solution for wavenumber} reduce the original set of six parameters now to only four independent parameters, and this further indicates that the four equations in \eqref{real phase variations} and \eqref{real phase variation 2} finally yield a closed modulation system.}
 
Equation \eqref{real phase variation 2} can be written as

\bse
\begin{align}
    \gamma_{T}-\left(-c_p^{2}\beta + Bc_p^{2}\right)_{X} &= 0,
    \label{second modulation equation}
    \\
    \beta_{T} + \gamma_{X} &= 0,
    \label{third modulation equation}
\end{align}
\ese
whereas Eq.~\eqref{real phase variations}, can be written as
\bse
\begin{align}
    \left(\frac{c_p}{12}\left(2W + c_pW_{c_p}\right) + \frac{2c_p}{K}\left(-\frac{1}{2}\beta^{2}+\frac{1}{2}C+\beta B\right)\right)_{T} \kern10em &
    \nonumber\\
    + \left(\frac{c_p^{2}}{12}\left(W_{c_p}c_p + W\right) + \frac{2c_p^{2}}{K}\left(-\frac{1}{2}\beta^{2}+\frac{1}{2}C+\beta B\right)\right)_{X} &= 0,\label{last me}\\
    K_{T} + \Omega_{X} &= 0.\label{cons of wa}
\end{align}
\ese

Taken together, the four equations in Eqs.~\eqref{second modulation equation}-\eqref{cons of wa}
form a closed system of modulation equations for the periodic solutions of the continuum model. One should also note that we expect that if the alternative strategy of averaging the conservation laws is applied, an equivalent Whitham modulation system will be obtained.
However, given that for the PDE model of interest the conservation laws are expressed at the level of displacements
(and not of strains), we do not pursue this 
avenue here.

\section{Riemann problems and DSW fitting}\label{s:riemann}

In this section, we discuss the setup of the Riemann problems for both the continuum PDE and the discrete granular DDE, as well as offer a comparison of the two. 

\subsection{Riemann invariants of the dispersionless system} 

Before discussing numerical simulations and the analytical method of DSW fitting, 
we  need to understand the dispersionless averaged version of the continuum model \eqref{Target PDE model}. First note that Eq.~\eqref{Target PDE model} can be written in the following form, 
\begin{equation}\label{eq: PDE in hydrodynamic form}
    \begin{aligned}
    \left(r - \frac{\epsilon^{2}}{12}r_{XX}\right)_{T} - \rho_{X} &= 0,\\
    \rho_{T} &= \left(r^{p}\right)_{X}.
    \end{aligned}
\end{equation}
According to \cite{El2005}, the Whitham modulation equations at both harmonic and solitonic edges should read
\bse
\label{eq: dispersionless version}
\begin{align}
     \left(\overline{r}\right)_T - \overline{\rho}_X &= 0,\\
    \overline{\rho}_T - \left[\left(\overline{r}\right)^{p}\right]_X &= 0.
\end{align}
\ese
These equations are obtained by setting the dispersion term to zero ($\epsilon=0$) in Eq.~\eqref{eq: PDE in hydrodynamic form} and averaging. 
We can then further put the first-order system \eqref{eq: dispersionless version} into the associated characteristic form which reads
\begin{subequations}
\label{eq: Riemann-invariates PDE}
\begin{align}
        \frac{\partial  q_1}{\partial T}+p^{\frac{1}{2}}\overline{r}^{\frac{p-1}{2}}\frac{\partial q_1}{\partial X} &= 0,\\
    \frac{\partial q_2}{\partial T} - p^{\frac{1}{2}}\overline{r}^{\frac{p-1}{2}}\frac{\partial q_2}{\partial X} &= 0,
\end{align}
\end{subequations}
where the two Riemann invariants are 
\begin{subequations}
\label{eq: def of two RIs}
    \begin{align}
        q_1 &= \overline{\rho} - \frac{2p^{\frac{1}{2}}}{p+1}\overline{r}^{\frac{p+1}{2}},\\
    q_2 &= \overline{\rho} + \frac{2p^{\frac{1}{2}}}{p+1}\overline{r}^{\frac{p+1}{2}}.
    \end{align}
\end{subequations}
which are associated with the two characteristic velocities $\lambda_{+} = p^{\frac{1}{2}}\overline{r}^{\frac{p-1}{2}}$ and $\lambda_{-} = -p^{\frac{1}{2}}\overline{r}^{\frac{p-1}{2}}$, respectively.

\subsection{Initial data for the continuum model}\label{IC discussion for rho}

The classic Riemann problem for the continuum model corresponds to the initial conditions
\begin{equation} \label{eq:step IC}
r(X,0) = \begin{cases} r^{-}\,, &  X< 0\,,\\ r^{+}\,, &  X>0\,, \end{cases} 
 \qquad
\rho(X,0) = \begin{cases} \rho^{-}\,, &  X< 0\,,\\ \rho^{+}\,, &  X>0\,, \end{cases} 
\end{equation}
These initial conditions are assumed when conducting the DSW fitting analysis
in Sec.~\ref{sec: dsw fitting}.
In what follows, we assume that $r_->r_+$.

The initial condition for the variable $\rho$ should satisfy a jump condition, as detailed in 
\cite{El2005}. In particular, this jump condition is obtained by demanding that the Riemann invariant $q_2$ (with associated characteristic velocity $\lambda_{-}$) of the dispersionless system Eq.~\eqref{eq: dispersionless version} to be constant. In other words, we must have that
\begin{equation}\label{eq:jc in implicit form}
    q_2\left(r^{+},\rho^{+}\right) = q_2\left(r^{-},\rho^{-}\right).
\end{equation}
Notice that the values of $\rho^{-}$ and $\rho^{+}$ are not known at this point, and we have to determine their values according to the jump condition specified in Eq.~\eqref{eq:jc in implicit form}. To this end, substituting the expression of $q_2$ defined in Eq.~\eqref{eq: def of two RIs} into the jump condition \eqref{eq:jc in implicit form}, we obtain that,
\begin{equation}\label{eq: jump condition}
    \rho^{-} + \frac{2\sqrt{p}}{p+1}(r^{-})^{\frac{p+1}{2}} = \rho^{+} + \frac{2\sqrt{p}}{p+1}(r^{+})^{\frac{p+1}{2}}.
\end{equation}
If considering step initial data, as defined in Eq.~\eqref{eq:IC}, one can freely select $r^+,r^-,\rho^+$
and then $\rho^-$ is determined via Eq.~\eqref{eq: jump condition}.

For numerical simulations, we will employ
a spectral method to discretize the spatial variables, and thus we require
initial conditions that will respect the periodic boundary conditions. The following ``box-type'' initial data 
are one such choice and are analogous to Eq.~\eqref{eq:IC}, 
\begin{equation}
r(X,0) = \begin{cases} r^{-}\,, & a<X<b\,,\\ r^{+}\,, & X<a~\vee~ X>b\,, \end{cases} 
 \qquad
 \rho(X,0) = \begin{cases} \rho^{-}\,, & a<X<b\,,\\ \rho^{+}\,, & X<a~\vee~ X>b\ \end{cases}
\label{eq:IC}
\end{equation}
where $a,b\in\mathbb{R}$ are two real constants with $a<b$. 
The discontinuity in this initial
data, however, makes it unsuitable for computations. Thus, for the numerical approximation of the PDEs, we employ a smooth approximation of the "box-type" initial strain,
\begin{equation}\label{IC for the strain}
    r(X,0) = r^{+} - \frac{r^{+} - r^{-}}{2}\left(\text{tanh}\left(50\left(X-a\right)\right) - \text{tanh}\left(50\left(X - b\right)\right)\right).
\end{equation}
We now need to find an appropriate smooth approximation of $\rho(X,0)$ that is consistent with Eq.~\eqref{eq:IC}
and that satisfies the jump conditions Eq.~\eqref{eq: jump condition}.
Assuming $r^+,r^-,\rho^+$ are fixed (for which
the nature of the equations allows the freedom
to arbitrarily select), and based on the jump condition of Eq.~\eqref{eq: jump condition}, we take the following initial condition for the variable $\rho$,
\begin{equation}\label{IC for rho}
    \rho(X,0) = \frac{2\sqrt{p}}{p+1}\left(\left(r^{+}\right)^{\frac{p+1}{2}} - r\left(X,0\right)^{\frac{p+1}{2}}\right) + \rho^{+}.
\end{equation}
This further suggests that the value of $\rho^{-}$ is given as follows,
\begin{equation}\label{eq:value of rho minus}
    \rho^{-} = \frac{2\sqrt{p}}{p+1}\left(\left(r^{+}\right)^{\frac{p+1}{2}} - \left(r^{-}\right)^{\frac{p+1}{2}}\right) + \rho^{+}.
\end{equation} 
which is clearly consonant with the jump condition Eq.~\eqref{eq: jump condition}. 

With $r_- > r_+$
and $\rho(X,0)$ chosen according to Eq.~\eqref{eq: jump condition}, a right-moving DSW
will form on the right of the "box", whereas a left-moving rarefaction wave
will form on the left of the "box". Since our focus is on the formation of
DSWs, we do not report results on the rarefaction waves.
See Secs.~\ref{s: first simulation} and \ref{s:numerics} for examples
of right-moving DSWs.

If one were to demand that the Riemann invariant $q_1$ be fixed (rather than
$q_2$) the jump condition Eq.~\eqref{eq: jump condition} would become
\begin{equation}\label{eq: jump condition2}
    \rho^{-} - \frac{2\sqrt{p}}{p+1}(r^{-})^{\frac{p+1}{2}} = \rho^{+} - \frac{2\sqrt{p}}{p+1}(r^{+})^{\frac{p+1}{2}}.
\end{equation}
It is worth noting that a choice of
$r_- > r_+$ and  $\rho(X,0)$ to satisfy this jump condition would lead to
a left-moving DSW on the left of the "box" and a  rarefaction wave
moving to the right. The results are qualitatively similar to the case
with $q_2$ fixed, and so this setting is not discussed further herein.

\subsection{Initial data for the discrete model}\label{s: IC DDE}

To set up the Riemann problem for the original granular DDE that is consistent with the initial data we just discussed
we need to further understand how the density variable $\rho$ relates to the discrete variables.
To this end, we consider an alternative first-order form of our continuum PDE which has a direct connection to the lattice system. In particular, we observe that instead of rewriting the continuum model as Eq.~\eqref{eq: PDE in hydrodynamic form}, it can also be written as follows,
\begin{subequations}
\label{eq: alternatice first-order form of the continuum model}
    \begin{align}
        r_T &= v_X, \\
        \left(v - \frac{\epsilon^{2}}{12}v_{XX}\right)_T &= \left(r^{p}\right)_X.
    \end{align}
\end{subequations}
Here, $r(X,T)$ still represents an approximation of the strain of the lattice $r_n$, but $v(X,T)$
is related to the particle velocity, $\dot{u}_n$. So now both variables of the PDE formulation are directly related to physical variables of the granular chain.
We then notice that, by setting $\rho = v - \frac{\epsilon^{2}}{12}v_{XX}$, the system \eqref{eq: alternatice first-order form of the continuum model} is equivalent to \eqref{eq: PDE in hydrodynamic form}. 
This gives an interpretation of $\rho$ in terms of the physical variables of the granular chain (namely in terms of the particle velocity).

\begin{figure}[t!]
    \centering
    \includegraphics[width=0.99\linewidth]{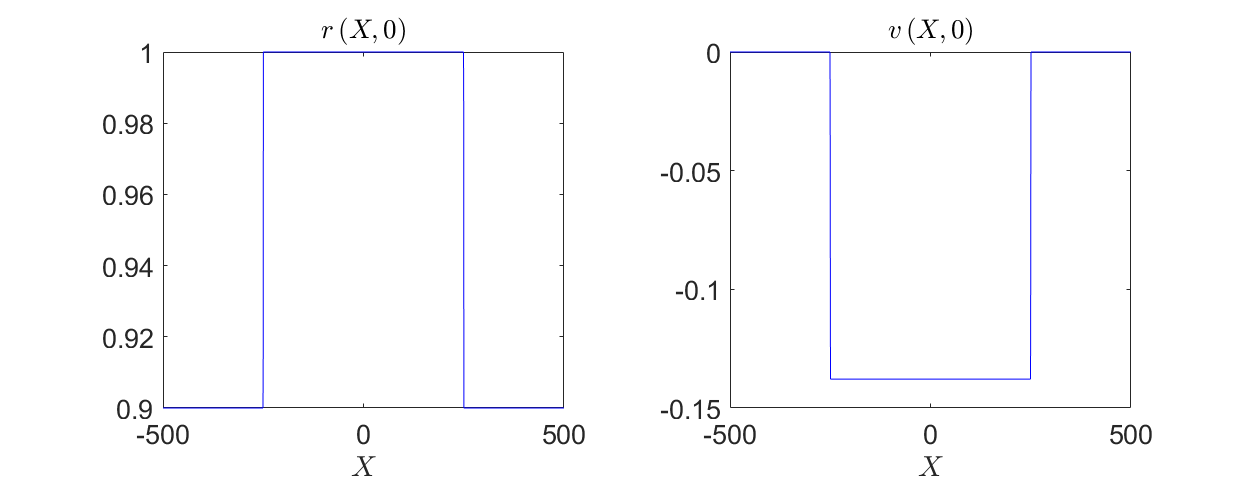}
    \caption{The initial conditions in Eqs.~\eqref{eq: ICs for the alternative first-order system}. 
    The background and parameter values are $r^{-} = 1$, $r^{+} = 0.9$, $p = 2$, and $\epsilon = 0.1$.}
    \label{fig4: ICs 7.14(a)-(b)}
\end{figure}

If we denote the initial conditions of Eq.~\eqref{eq: PDE in hydrodynamic form} as
\begin{subequations}
\label{eq: ICs for the alternative first-order system}
    \begin{align}
        r\left(X,0\right) &= f(X),\\
        \rho\left(X,0\right) &= g(X)
    \end{align}
\end{subequations}
then the associated initial conditions for the system \eqref{eq: alternatice first-order form of the continuum model} read
\begin{subequations}
\label{eq: ICs for the alternative first-order system}
    \begin{align}
        r\left(X,0\right) &= f(X),\\
        v\left(X,0\right) &= \mathcal{F}^{-1}\left[\frac{\mathcal{F}\left[g(X)\right]}{1 + \frac1{12}{\epsilon^{2}K^{2}}}\right]
         = \int_{-\infty}^\infty h(X')g(X - X')\,dX'\,,
    \end{align}
\end{subequations}
where $\mathcal{F}^{-1}, \mathcal{F}$ refer to the inverse and usual Fourier operators, $K$ the Fourier wave number, 
\begingroup
and 
\begin{equation}
h(X) = \mathcal{F}^{-1}\bigg[\frac1{1 + \frac1{12}{\epsilon^{2}K^{2}}}\bigg] = (\sqrt3/\epsilon)\,e^{-2\sqrt3|X|/\epsilon}\,.
\end{equation}
Figure~\ref{fig4: ICs 7.14(a)-(b)} showcases the profiles of the initial conditions in Eq.~\eqref{eq: ICs for the alternative first-order system}.
\endgroup

Now, we finally consider the Riemann problem of the granular DDE. First, we rewrite the second-order discrete system \eqref{e:Strain ODE} as the following equivalent first-order discrete system,
\begin{subequations}
\label{eq: first-order equivalent discrete system}
    \begin{align}
        \dot r_n &= s_n,\\
        \dot s_n &= r_{n+1}^{p} - 2r_n^{p} + r_{n-1}^{p},
    \end{align}
\end{subequations}
where $s_n$ is the strain derivative. Similar to the initial conditions in \eqref{eq: ICs for the alternative first-order system}, the Riemann initial conditions for the discrete system \eqref{eq: first-order equivalent discrete system} read
\begin{subequations}
\label{eq: Riemann ICs for DDEs}
    \begin{align}
    r_n(0) &= f\left(n\epsilon\right),\\
    s_n(0) &= v\left(n\epsilon, 0\right) - v\left(\left(n-1\right)\epsilon, 0\right),
    \end{align}
\end{subequations}
where the right-hand side $v$ in Eq.~\eqref{eq: Riemann ICs for DDEs} comes from the continuum ICs expressed in \eqref{eq: ICs for the alternative first-order system}(b).

\subsection{Numerical simulation and DSWs} \label{s: first simulation}

\begin{figure}[t!]
    \centering
    \includegraphics[width=0.99\linewidth]{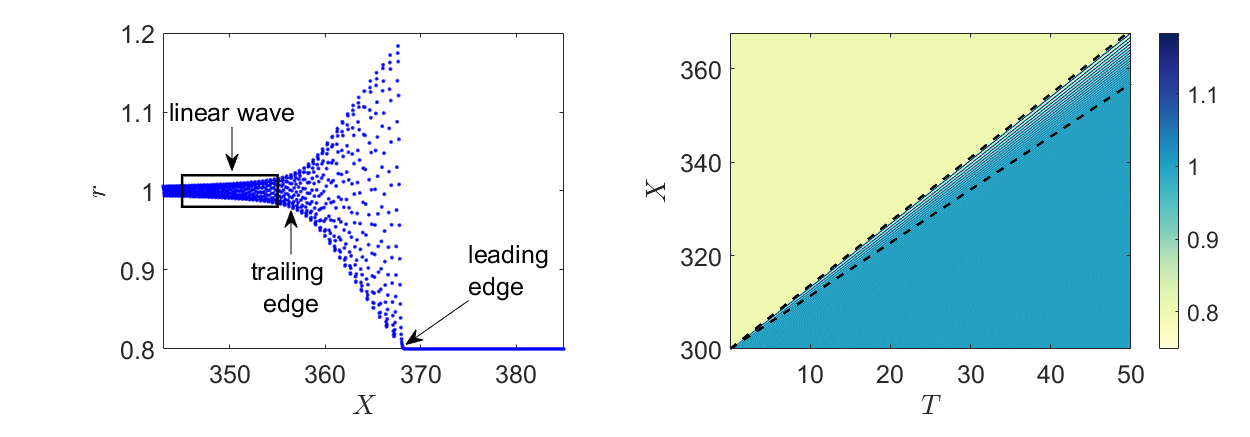}
    \caption{Numerical simulation of the Riemann problem: The left panel shows the profile of the DSW of the continuum model \eqref{Target PDE model} at $T = 50$, while the right panel depicts the plot of the strain magnitude $r(X,T)$, and the two black dashed lines in the density plot represent the leading (upper) and the trailing (lower) edges of the DSW, respectively. In particular, the upper and the lower dashed black lines depict $X = s^{+}T$ and $X = s^{-}T$ where $s^{+}, s^{-}$ are obtained based on Eqs.~\eqref{eq: leading and trailing-edge speeds}. Meanwhile, notice that the parameters of $r^{+}, r^{-}$ in the Riemann initial condition \eqref{eq:IC} are set to be $r^{+} = 0.8$, and $r^{-} = 1$ in the numerical simulation.}
    \label{fig: DSW contour and profile plots}
\end{figure}

Figure~\ref{fig: DSW contour and profile plots} displays the simulation of the Riemann problem for the DDE. A waveform emerges at the right upper corner of the "box-type" initial data.  
This waveform connects left state ($r^-$) to the right state
($r^+$) and expands as time increases. If one zooms
into a small spatial-window of the waveform, it will appear
as though the wave is periodic. The underlying parameters
of the periodic wave will depend on where in the lattice
you zoom, and hence the waveform is modulated. This is
the so-called DSW. At the front of the wave (called the leading
edge) the wave closely resembles a solitary wave,
and it travels with a near constant speed $s^+$ and has some
amplitude $a^+$.
On the other hand, the back part of the DSW (called the trailing edge) is characterized by a linear wave with some wavenumber $k^-$ traveling at a speed $s^{-}$ with amplitude that vanishes
as the trailing edge is approached from within the DSW.
For any finite time snapshot of the DSW, there will be 
linear waves that are located around the background value of $r^{-}$. These linear waves are not a part of the DSW per se,
and vanish for $t\rightarrow \infty$.
To avoid any possible confusion, we point out that the PDE~\eqref{Target PDE model} is non-dispersive only for small-amplitude waves on a zero background.
Conversely, it is fairly straightforward to see that the PDE does admit small-amplitude waves 
riding on top of any non-zero background, which is what we refer to with the term linear waves.
Figure~\ref{fig: DSW contour and profile plots}(a) shows a time snapshot of the
DSW, where the key elements of the DSW, namely the leading
and trailing edge can be seen, along with the linear waves
that are always present for any finite time snapshot.
Figure~\ref{fig: DSW contour and profile plots}(b)
shows a plot of the strain magnitude $r(X,T)$
where the expanding oscillatory region (i.e., the DSW) can be seen.
The dashed lines in panel (b) are analytical approximations
of the leading and trailing edges that stem from
the so-called DSW fitting method. Before discussing
those approximations, we discuss some preliminaries
involving the the dispersionless system.

\subsection{DSW fitting} \label{sec: dsw fitting}

Next, we consider special reductions of Whitham's modulation system that we have derived in Sec.~\ref{s:whitham}.
These reductions will be used to 
obtain valuable information about the dispersive shock waves, such as the leading and trailing edge speeds
as well as the amplitude of the solution at the leading edge of the DSW and the wavenumber at its trailing edge.
To this end, we apply the well established,
and highly successful in the continuum limit
``DSW fitting" method of El~\cite{El2005},
which has not been much explored in the lattice setting.

We note that the applicability of the method requires the underlying modulation equations to be
strictly hyperbolic and genuinly nonlinear, e.g., see 
\cite{El2005,PhysRevE.88.023016,Hoefer2014}.
It is not known at present whether the modulation equations derived in Section~\ref{s:whitham}
satisfy these requirements.
On the other hand, we will proceed under the assumption that these conditions are satisfied,
and we will validate the calculations a posteriori by comparing the theoretical predictions 
with the results of direct numerical simulations of both the PDE~\eqref{Target PDE model}
and the original DDE~\eqref{e:Strain ODE}, demonstrating good agreement.

One preliminary and important task is to compute the associated linear dispersion relation of the original continuum model \eqref{Target PDE model}.
Leveraging our earlier analysis in Eq.~\eqref{e:NLWPDEO2disprel},
we recall that the linear dispersion relation reads:
\begin{equation}
\Omega_0^2(A,K) = {pA^{p-1}K^2} \bigg/ \bigg( 1+\frac{\epsilon^2}{12}K^2 \bigg)\,.
\label{ldr}
\end{equation}
Then, in line with the general DSW fitting theory of~\cite{El2005},
the trailing-edge wave number $K^{-}$ and the leading-edge conjugate wave number $\widetilde{K}^{+}$ have to satisfy the following two boundary value problems,
\bse
\label{eq: two BC problems}
\begin{align}
        \frac{dK}{d\overline{r}} &= \frac{\partial\Omega_0/\partial\overline{r}}{\lambda_{+}\left(\overline{r}\right) - \partial\Omega_0/\partial K}, \hspace{5mm} K(r^{+}) = 0,\\
            \frac{d\widetilde{K}}{d\overline{r}} &= \frac{\partial \widetilde{\Omega}_s/\partial\overline{r}}{\lambda_{+}\left(\overline{r}\right) - \partial\widetilde{\Omega}_s/\partial\widetilde{K}}, \hspace{5mm} \widetilde{K}(r^{-}) = 0,
\end{align}
\ese
where $r^{+}, r^{-}$ refer to the two background values
of the strain in the Riemann initial condition (defined in Eq.~\eqref{eq:IC}) representing
a jump from $r^{-}$ to $r^{+}$.
We further notice that the notation $\widetilde{\Omega}_s$ 
represents the conjugate linear dispersion relation which is determined according to the prescription of~\cite{El2005} as:
\begin{equation}\label{eq: conjugate dispersion relation}
    \begin{aligned}
        \widetilde{\Omega}_s(\overline{r},\widetilde{K}) = -i\Omega_0(\overline{r},i\widetilde{K}),
    \end{aligned}
\end{equation}
where the linear dispersion relation $\Omega_0$ is given in \eqref{ldr}.

A direct integration of both ODEs in \eqref{eq: two BC problems} with the help of the boundary conditions yields the following two transcendental equations which are used to determine the numerical values of trailing-edge wavenumber and of 
the leading edge conjugate wavenumber,

\begin{equation}
\label{eq: two transcendental equations}
\begin{aligned}
        \log\left|\frac{\overline{r}}{r^{+}}\right| = \frac{2}{p-1}&\left(\sqrt{1+\frac{1}{12}\epsilon^{2}K^{2}} - \frac{1}{2}\log\left|1 + \frac{1}{2}\sqrt{4 + \frac{\epsilon^{2}K^{2}}{3}}\right| + \right. \\
        &\left. \frac{1}{2}\log\left|\frac{4\left(1-\frac{1}{2}\sqrt{4+\frac{\epsilon^{2}K^{2}}{3}}\right)}{\epsilon^{2}K^{2}}\right| + \frac{1}{2}\log\left(12 + \epsilon^{2}K^{2}\right) - 1\right)
        \\\log\left|\frac{\overline{r}}{r^{-}}\right| = \frac{2}{p-1}&\left(\sqrt{1-\frac{1}{12}\epsilon^{2}\widetilde{K}^{2}} - \frac{1}{2}\log\left|1+\frac{1}{2}\sqrt{4-\frac{\epsilon^{2}\widetilde{K}^{2}}{3}}\right|+\right.\\
        &\left.\frac{1}{2}\log\left|12-\epsilon^{2}\widetilde{K}^{2}\right|+\frac{1}{2}\log\left|\frac{4\left(1-\frac{1}{2}\sqrt{4-\frac{\epsilon^{2}\widetilde{K}^{2}}{3}}\right)}{\epsilon^{2}\widetilde{K}^{2}}\right| - 1\right).
    \end{aligned}
\end{equation}

To compute numerically the trailing-edge wavenumber and the leading-edge conjugate wavenumbers, we first notice that all the parameters in Eqs.\eqref{eq: two transcendental equations} are known except $K$ and $\widetilde{K}$ which are the target unknown variables to be sought. 
To obtain the latter, we observe that at the trailing and leading edges, 
$\overline{r} =  r^{-}$ and $\overline{r} = r^{+}$, respectively. We then substitute all the known parameters in Eqs.\eqref{eq: two transcendental equations} to solve for $K$ and $\widetilde{K}$.

In addition, the trailing and leading edge speeds, denoted by $s^{-}$ and $s^{+}$, respectively, can be computed using the following formulas of the DSW fitting method:
\begin{subequations}
\label{eq: leading and trailing-edge speeds}
    \begin{align}
        s^{-} &= \frac{\partial\Omega_0}{\partial K}\left(r^{-}, K^{-}\right) = \frac{p^{\frac{1}{2}}\left(r^{-}\right)^{\frac{p-1}{2}}}{\left(1 + \frac{1}{12}\left(\epsilon K^{-}\right)^{2}\right)^{\frac{3}{2}}},\\
        s^{+} &= \frac{\widetilde{\Omega}_s}{\widetilde{K}}\left(r^{+}, \widetilde{K}^{+}\right) = \frac{p^{\frac{1}{2}}\left(r^{+}\right)^{\frac{p-1}{2}}}{\left(1 - \frac{1}{12}\left(\epsilon\widetilde{K}^{+}\right)^{2}\right)^{\frac{1}{2}}}.
    \end{align}
\end{subequations}

An important observation here is that in expression~\eqref{eq: two transcendental equations} and \eqref{eq: leading and trailing-edge speeds},
there is only a single effective parameter $\epsilon K$. This implies that once we return to the original
granular variables and compare the wavenumber prediction of the DSW fitting and the wavenumber
corresponding to the lattice, the prediction once again will be independent of $\epsilon$,
in light of the relationship $k = \epsilon K$. The same is true for the leading and trailing
edge speeds, namely they are independent of $\epsilon$, much like the case of the solitary
waves, whose predictions were also independent of $\epsilon$. Finally, we recall that we can also compute the analytical prediction for the soliton amplitude for the cases of $p = 2, 3$ by using the formulas in Eq.~\eqref{eq: explicit soliton amplitude formula for the case p = 2} and Eq.~\eqref{Explicit formula for soliton amplitude for p = 3}, respectively, with replacing $c$ by $s^{+}$ in Eqs.~\eqref{eq: leading and trailing-edge speeds}.

\section{Numerical validation}\label{s:numerics}

We are now ready to validate the theoretical results obtained previously by numerically solving the Riemann problems and comparing the numerical results with the associated quantities derived from the DSW fitting method. We employed a fourth-order Runge-Kutta method for time stepping both the discrete model \eqref{e:Strain ODE} and the quasi-continuum PDE \eqref{Target PDE model},
as well as pseudo-spectral methods to discretize the spatial derivatives in Eq.~\eqref{Target PDE model}.

Firstly we present the DSW profiles comparison as shown in the following figures for the cases of $p = 3$ and $p = 2$, respectively. Note that the smallness parameter $\epsilon$ is set to be $0.1$ over all PDE simulations. 
To make a direct comparison between the DSWs obtained from the PDE and DDE simulations, when the spatial variable of the PDE simulation is taken to be $X$, we consistently define the DDE spatial variable to be $n\epsilon$ 
per the principal relationship between the PDE and DDE spatial domains $X = n\epsilon$.

\begin{figure}[b!]
\begin{subfigure}[h]{0.45\linewidth}
\includegraphics[width=\linewidth]{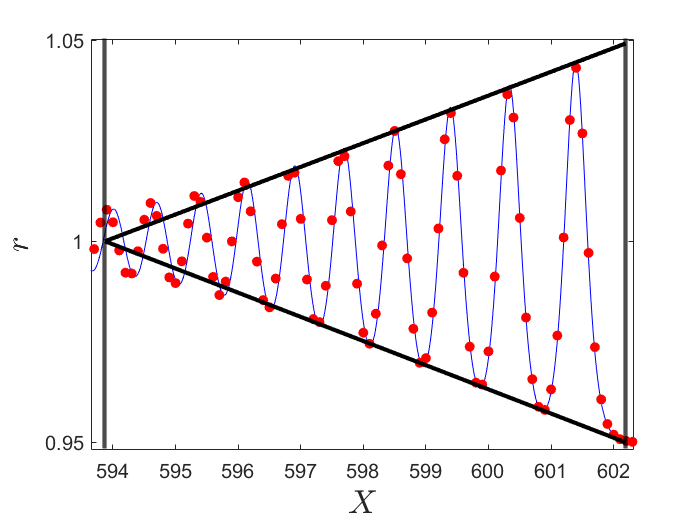}
\end{subfigure}
\hfill
\begin{subfigure}[h]{0.45\linewidth}
\includegraphics[width=\linewidth]{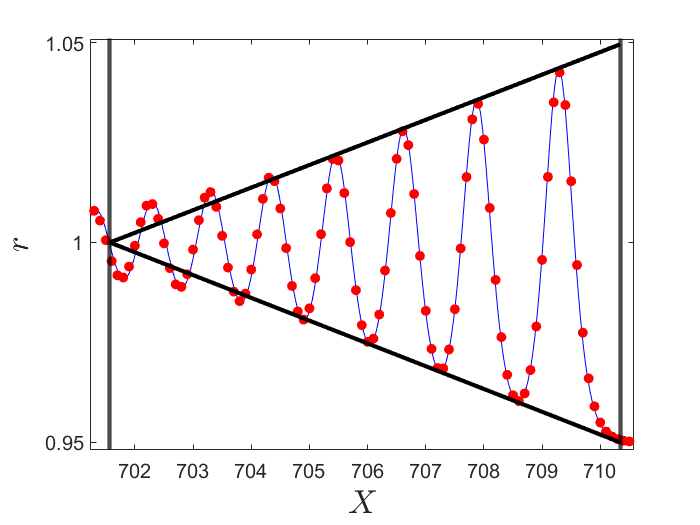}
\end{subfigure}%
\caption{Comparison of the DSW profiles for the cases of $p = 3$ and $p = 2$: in both left and right panels, the blue solid curves represent the DSW simulated from the regularized
continuum PDE model \eqref{Target PDE model}, and the red dots refer to the DSW of the DDE simulation. The black triangular envelope denotes the theoretical DSW fitting. Notice that the initial condition for the strain variable $r$ is given in Eq.~\eqref{IC for the strain} where $a = 200, b = 500$, and the background information is given by $r^{+} = 0.95, r^{-} = 1$. The corresponding initial condition for the density $\rho$ is given in Eq.~\eqref{IC for rho}, and the computational domain is $X \in \left[0, 1000\right]$. Notice that the left panel refers to the evolution at $T = 60$ (for the PDE \eqref{Target PDE model}) and $t = 600$ (for the DDE \eqref{e:Strain ODE}), while the right panel denotes the evolution dynamics at $T = 150$ (for the PDE \eqref{Target PDE model}) and $t = 1500$ (for the DDE \eqref{e:Strain ODE}).}
\label{figure: DSW profile comparison}
\end{figure}

Figure~\ref{figure: DSW profile comparison} shows the granular
lattice subject to the Riemann initial data given by Eq.~\eqref{eq: Riemann ICs for DDEs}
with $r^-=1$ and $r^+=0.95$ (see the red dots). Only a zoom of the right-moving wave is shown,
which is where the DSW appears. The solid blue lines are the numerical simulation of the regularized
continuum PDE proposed herein, namely Eq.~\eqref{Target PDE model} with
the consistent initial data given by Eqs.~\eqref{IC for the strain}
and ~\eqref{IC for rho}. The lattice and PDE simulations agree very well.
The solid black lines represent the approximations obtained from the DSW fitting method
described previously. In particular for panel (a), the vertical lines represent the trailing and leading edges, namely, $60 s^-$ and $60 s^+$, respectively, where $t=60$ is the time chosen and $s^-,s^+$ are the speeds from Eq.~\eqref{eq: leading and trailing-edge speeds}.
The sloped lines are an approximation of the envelope, and were obtained by connecting
the point $(60 s^-,r^-)$ with $(60 s^+, a^+ + r^+)$ (the top line) and the line
defined by the points $(60 s^-,r^-)$ with $(60 s^+, r^+)$, (the bottom line).
Panel (b) is similar but the time $t=150$ is used in place of $t=60$.
The lattice simulations are encompassed by the triangle enclosed by the solid
lines demonstrating that the theoretical prediction of the DSW characteristics (based on the DSW fitting
discussed above) agrees quite well with the numerical observations. 

Such a quantitatively adequate fitting 
(everywhere but for part of the DSW's linear tail)
also indicates that the proposed PDE model~\eqref{Target PDE model} succeeds in characterizing the DSW embedded in the discrete model. 
The inaccuracy at the tail is common when approximating DSWs \cite{El_2017}.
While the linear tails decay in time (with the rate $t^{-1/3}$) \cite{MielkePatz2017}
it will always be present in any finite simulation.

As an aside, it is worth 
mentioning that we have also 
simulated the models of~\cite{Ahnert_2009}
(i.e., Eq.~(\ref{not good model}))
and~\cite{Nester2001} (i.e., Eq.~(\ref{Pikovsky's continuum model})) for the 
Riemann problem
initial conditions of the present work.
In line with the expectation associated
with their dispersion relation discussed
earlier, we have found that while these
models form a DSW, the latter keeps increasing
in amplitude indefinitely as a result of
the ultraviolet catastrophes present in the models.
This corroborates our expectation that such
models, while meaningful towards a traveling or
periodic
wave analysis, cannot be used for the 
initial value problem considered herein.

To further quantify the  fitting of the DSWs, especially
in connection to the theoretical predictions, we compute numerically some important quantities relevant to the DSWs including (i) leading-edge speeds, (ii) leading-edge soliton amplitudes, (iii) trailing-edge speeds, and (iv) trailing-edge 
wavenumbers. 
Before we display the numerical results of the four quantities, we first discuss the method applied to measure these quantities in the numerical simulation. For the leading-edge calculations including items (i) and (ii), we pinpoint the location of the leading-edge as the $x$ coordinate of the highest peak of the DSW. We then compute the leading-edge speed by first recording the leading-edge locations 
at various times $t$. From this set of data we find the best-fit line
and use the corresponding slope as the prediction for the leading-edge speed. On the other hand, for the numerical leading-edge soliton amplitudes denoted by $a^{+}$, it can be computed numerically via the following formula,
\begin{equation}\label{eq: numerical soliton amplitude formula}
    a^{+} = \max_{X}\{u\left(X,T_f\right)\} - r^{+},
\end{equation}
where $u\left(X,T_f\right)$ is the numerical solution to the PDE/DDE models at the final time of the simulation $t = T_f$. Moreover, for the theoretical leading-edge soliton amplitudes, we compute them by applying the formulas \eqref{eq: explicit soliton amplitude formula for the case p = 2} and \eqref{Explicit formula for soliton amplitude for p = 3} for the cases of $p = 2$ and $p = 3$, respectively.
Next, for the trailing-edge computation, to identify the trailing-edge location needed for (iii) and (iv),
we find the best fit line passing through a set of local maxima 
(which set is described below) and a best fit line passing through a set of 
local minima. Similar to the sloped lines of Fig.~\ref{figure: DSW profile comparison},
these two lines will intersect. That intersection point will act as measurement
of the numerical trailing edge. In order to find a suitable interval
of maxima (and minima) we first define the quantities
\vspace*{-1ex}
\begin{subequations}
\label{eq: core of the DSWs}
    \begin{align}
    a^{u} &= r^{-} + \frac{\left|r^{-} - r^{+}\right|}{N},\\
    a^{l} &= r^{-} - \frac{\left|r^{-} - r^{+}\right|}{N},
    \end{align}
\end{subequations}
where $N\in\mathbb{Z}$ is a positive integer.
These values are approximately $3/4$ and $1/4$ the amplitude of the DSW, respectively.
Then, the sets of local maxima and
minima are those with values that fall in the intervals, respectively,  
\begin{subequations}
\label{neightborhood}
    \begin{align}
        N^{u} &= \left(a^{u} - \nu, a^{u} + \nu\right),\\
        N^{l} &= \left(a^{l} - \nu, a^{l} + \nu\right).
    \end{align}
\end{subequations}
where $\nu > 0$ is a small constant which can be defined as $\nu = \frac{\left|r^{-} - r^{+}\right|}{5}$. We notice that the number $\nu$ is not always fixed as the value of $r^{+}$ gets closer to that of $r^{-}$ we expect less oscillations to occur in the core of the DSW and hence we need a larger $\nu$ as $r^{+}$ becomes greater.
Finally, we find the best fit lines going through the peaks in these
intervals and treat the intersection measured trailing-edge location
(for both the DDE and PDE simulations).

\begin{figure}[t!]
\begin{subfigure}[h]{0.4\linewidth}
\includegraphics[width=\linewidth]{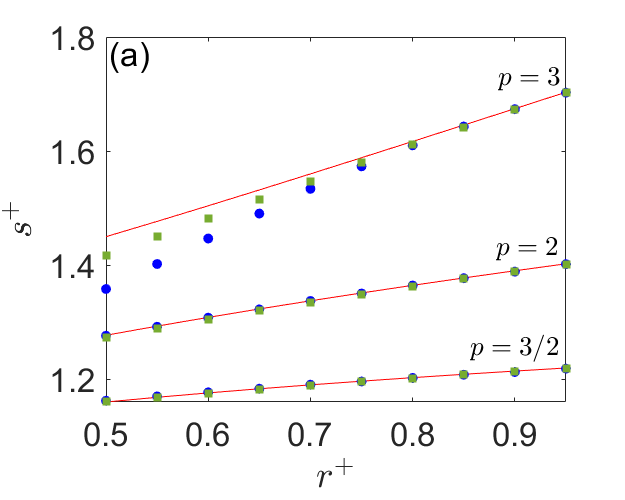}
\end{subfigure}
\hfill
\begin{subfigure}[h]{0.4\linewidth}
\includegraphics[width=\linewidth]{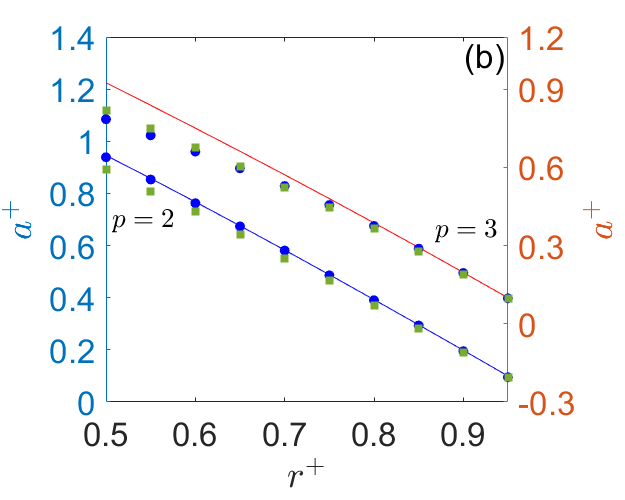}
\end{subfigure}%
\caption{The leading-edge quantities comparison: the left and the right panel refer to the comparison of the leading-edge speeds and soliton amplitude, respectively. The red solid line represents the theoretical prediction of the leading-edge quantities, while the blue circles and green squares refer to the numerical leading-edge quantities obtained from the simulation of the regularlized continuum PDE and the lattice DDE, respectively. The background information $r^{-} = 1$ is fixed, but $r^{+}$ is varied. Here $\epsilon = 0.1$.}
\label{figure: numerical leading-edge quantities comparison}
\end{figure}

We display first the comparison of the leading-edge speed in Fig.~\ref{figure: numerical leading-edge quantities comparison}(a) for $r^-=1$ fixed and various
$r^+$.  The value $\epsilon=0.1$ is fixed (although recall that the prediction of the DSW
fitting method is independent of $\epsilon$). The red line indicates the theoretical prediction from the DSW fitting method, see Eqs.~\eqref{eq: leading and trailing-edge speeds}, where 
the blue circles and green squares are the measured speeds
from the numerical solution of the PDE and DDE, respectively. 
This is done for three values of the nonlinearity $p=3/2,2,3$. In all three
cases, the agreement is very good overall. As the jump height
$|r^+ - r^-|$ decreases, the agreement between the DSW fitting
prediction and simulation becomes better. For all values of $r^+$ considered,
the blue dots fall almost perfectly into the green squares,
indicating the strong agreement between DDE and PDE simulations.
Figure~\ref{figure: numerical leading-edge quantities comparison}(b)
is similar to panel (a), but now the leading-edge amplitude is shown.
Once again the markers are from the DDE and PDE simulations. 
Here, we only show results for $p=2$ and $p=3$, since we do not have
an explicit formula for the amplitude in the case $p=3/2$.
The leading-edge amplitude formulas are from
Eq.~\eqref{eq: explicit soliton amplitude formula for the case p = 2}
for $p=2$ and Eq.~\eqref{Explicit formula for soliton amplitude for p = 3}
for $p=3$ where the speed $c$ is replaced by $s^+$ in the formulas.

While the quantitative agreement is not as good for larger jump heights
when compared to panel (a), the overall trends are similar.
Considering both panels (a,b) together, we can see that as the jump height
decreases, the DSW increases its speed, but its amplitude decreases.
While this may seem initially counter-intuitive (based on the situation
for solitary waves), it is actually quite natural, since the background
 of the leading edge solitary wave is increasing (as the jump height
 decreases).

\begin{figure}[b!]
    \centering
    \includegraphics[width=1.05\linewidth]{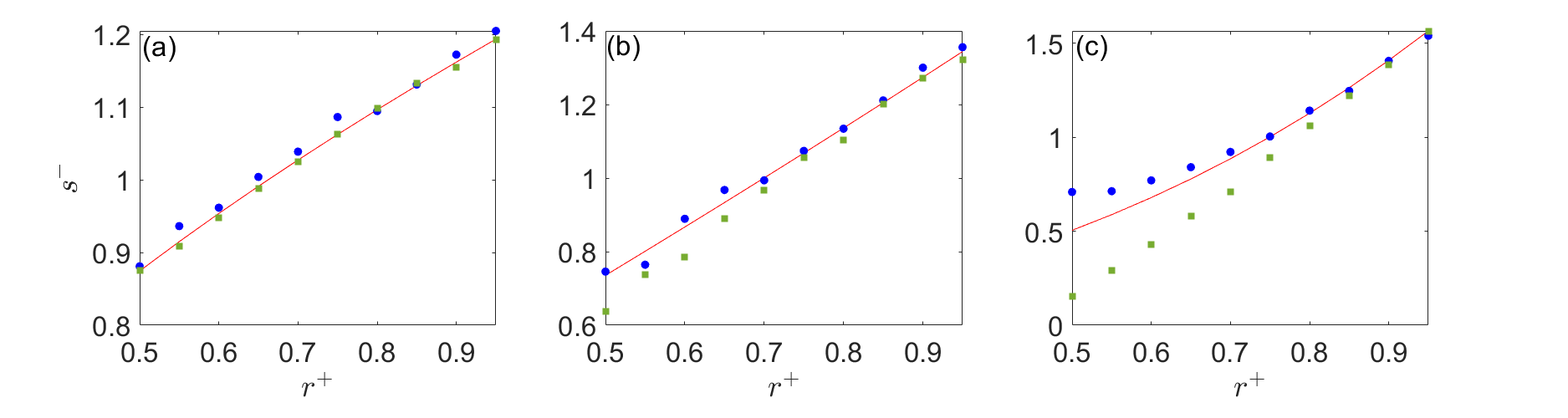}
    \caption{The trailing-edge speeds comparison: Note that the panels (a), (b), and (c) refer to the leading-edge speeds comparison for the cases of $p = 3/2, 2, 3$, respectively. In addition, the solid red curves depict the analytical prediction of the trailing-edge speeds based on the formula of $s^{-}$ in Eqs.~\eqref{eq: leading and trailing-edge speeds}, while the blue circles and green squares showcase the numerically measured trailing-edge speeds of the continuum PDE \eqref{Target PDE model} and the associated DDE \eqref{e:Strain ODE}, respectively.}
    \label{figure: trailing-edge speeds comparison}
\end{figure}

\begin{figure}[t!]
    \centering
    \includegraphics[width=0.35\linewidth]{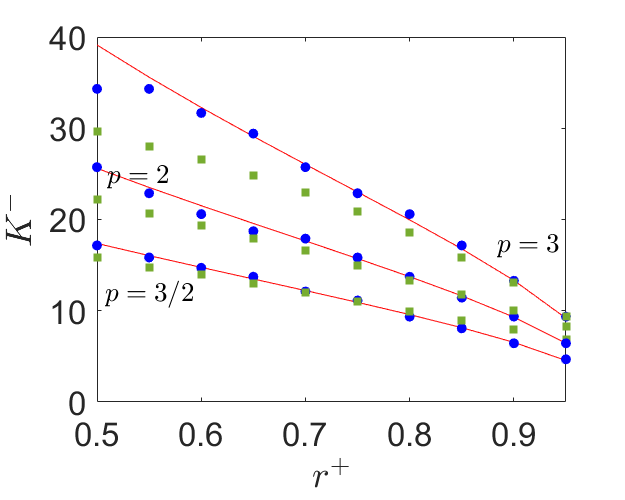}
    \caption{The trailing-edge wavenumbers comparison: Notice that the red solid curves, from the top to the bottom, depict the analytical prediction of the trailing-edge wavenumbers, while the blue circles and green squares refer to the numerically measured trailing-edge wavenumbers of the continuum PDE model \eqref{Target PDE model} and the corresponding discrete granular model \eqref{e:Strain ODE}.}
    \label{figure: trailing-edge wavenumbers comparison}
\end{figure}

Next, we investigate the trailing-edge quantities comparison, shown
in Figs.~\ref{figure: trailing-edge speeds comparison}-\ref{figure: trailing-edge wavenumbers comparison}.
In Fig.~\ref{figure: trailing-edge speeds comparison} the trailing edge speed is shown,
where the formula used to obtain the trailing edge speed
is from Eq.~\eqref{eq: leading and trailing-edge speeds}.
We notice that the trailing edge speeds of the DSWs from the numerical simulation of the Riemann problems are obtained by estimating the slope of $x$-$t$ plane where $x$ denotes the coordinate of the trailing edge locations measured by the method described above, and $t$ the temporal coordinate which corresponds to the time that the numerical solution is saved. The same marker and line conventions are used as in the previous figure.
First, we notice clearly that the agreement is not as favorable as the one we have seen for the leading edge.  As discussed above, there is always the presence of a linear tail
in the finite simulations, and this makes the estimation of the trailing
edge more prone to estimate errors. Thus it is possible that the larger
deviations between the numerical simulations and the DSW fitting predictions
are due to the method used to estimate the trailing edge in the simulation,
rather than to inaccuracies in the DSW fitting approach. 
Nevertheless, as we can observe from both Figs.~\ref{figure: trailing-edge speeds comparison}-\ref{figure: trailing-edge wavenumbers comparison},
the theoretical and numerical computed quantities still agree quite reasonably in the sense that all of them are rather proximal and indeed 
do not deviate significantly from one another. 
Moreover, no signatures of non-monotonicity,
e.g., of the trailing edge speed on the jump height
are observed here (a feature that has been
observed elsewhere, such as, e.g.,~\cite{ElGrimshawSmyth2006}).
Figure~\ref{figure: trailing-edge wavenumbers comparison}
is similar to Fig.~\ref{figure: trailing-edge speeds comparison}, but now the wavenumber is shown. 
The formula used to obtain the trailing edge wavenumber
is obtained by solving Eq.~\eqref{eq: two transcendental equations}.
To compute the wavenumber in the simulations, we first compute
the inverse of the phase speed by comparing the time series of
two consecutive nodes near the trailing edge. The inverse phase speed is then multiplied by the frequency to give an approximation of the wavenumber.

Notice that we have also conducted the numerical experiment with the initial condition \eqref{IC for the strain} with the number $50$ replaced by $10$ and we do not find this change has made an appreciable effect on the numerically measured trailing-edge wavenumber.

Finally, it is also worthwhile to mention that we need to be careful about the fact that $K = \epsilon^{-1} k$ which shows how the wavenumbers of the continuum PDE \eqref{Target PDE model} and the discrete granular model \eqref{e:Strain ODE} are related. Hence, in the numerical simulation of the granular discrete system \eqref{e:Strain ODE}, we scale and use the computational domain  $X = \epsilon n$, where $n$ denotes the lattice site, so that the estimated wavenumber in the discrete system can be compared to that of the continuum model.

The overall agreement between the various wavenumbers is quite good, despite larger deviations, stemming, possibly, from the uncertainty in the numerical wavenumber estimation.
Considering both panels (a,b) together, we can see that as the jump height
decreases, the trailing edge speed increases (just like the leading edge speed)
and the wavenumber decreases. It is interesting to note that, if we 
consider
the wavenumber in terms of the granular variables (where the value of
the wavenumber will be multiplied by $\epsilon=0.1$) all wavenumbers are below
$\pi$. The latter would correspond to a binary oscillation in the lattice. While
the wavenumbers are bounded in the discrete problem, namely, being confined
to $[0,\pi]$, the wavenumber is unbounded in the PDE. Indeed, it has been
conjectured in \cite{Venakides99} (see also \cite{Venakides91} which includes
a rigorous treatment in the special case of the Toda lattice) that the wavenumber reaching $\pi$ within the core of the DSW
represents a key change in the structure of the DSW. 
In that case,
the DSW would connect to a periodic (binary) wave, rather than being connected
to a constant. Such a feature could not be captured by the PDE model, and highlights
the fact that larger errors are to be expected for larger wavenumbers and (according
to Fig.~\ref{figure: trailing-edge wavenumbers comparison}) larger jump heights.

The very good agreement of the DSW fitting method seen
in the previous two figures demonstrates that,
while it is not straightforward
to solve the Whitham modulation equations (even though these
equations can indeed be written explicitly for the PDE model),
the leading and trailing edge analysis still yield important information regarding
characteristics of the DSW. This is revealed
indirectly via the DSW fitting method, which is built upon
the modulation theory, but never explicitly uses the latter.
Indeed, therein lies the power of the approach
derived in \cite{El2005}. While this approach
has been employed in several continuum contexts,
the results shown here show that it is successful
in the DDE context as well. In particular,
one can identify
the key features (speed, amplitude) of the leading edge,
as well as those (speed, wavenumber) of the trailing edge
and the self similar nature of the pattern in between allows
for a complete characterization thereof in the form of a 
modulated periodic wave between the two limits.

\section{Conclusions and future challenges}\label{s:conclusions}

In the present work we have revisited the topic of
dispersive shock waves in the lattice nonlinear dynamical
system setting of granular crystals, i.e., a platform
of wide, ongoing theoretical, numerical and experimental interest.
Along the way, we proposed a novel ---to the best of our knowledge---
regularized continuum (PDE) approximation of the lattice
model. This led us to explore the prototypical
workhorse of this nonlinear model, namely its
traveling wave structures and to compare the discrete 
numerically exact form thereof with the different
continuum approximations that have been formulated
in the literature. En route to describing the DSWs,
we have also identified the periodic (cnoidal wave)
solutions of the model. We have also examined
the conservation laws of the regularized continuum
approximation, as well as formulated based on the
Lagrangian description (one of the possible 
constructions thereof) the Whitham modulation equations
for this model. While these equations do not appear
to us amenable to straightforward analysis at the
present stage (an obstacle often encountered in
non-integrable systems), we used as a way to bypass
this difficulty the application of the DSW fitting method,
through an early, prototypical example of such an application
in lattice nonlinear dynamical systems. We have found the
latter method to work very well in suitable regimes offering
a (not only qualitatively, but also semi-quantitatively)
satisfactory description of both the leading and trailing
edges, and through those of the envelope and the self-similar
structure of the DSW of both the regularized PDE and the
lattice DDE, which were also in good agreement with each other.
This was illustrated through systematic numerical simulations
comparing all three of the above DSWs (theoretical one based
on the DSW fitting analysis of the continuum model, continuum model simulation and lattice
model dynamics).

It is relevant to once again point out that the applicability of the DSW fitting method requires
the corresponding modulation equations be genuinely nonlinear and strictly hyperbolic,
e.g., see \cite{El2005,ElGrimshawSmyth2006,PhysRevE.88.023016,Hoefer2014},
and it is not known at present whether the modulation equations derived in section~\ref{s:whitham} enjoy these properties.
When the modulation equations are not genuinely nonlinear and strictly hyperbolic, the trailing edge speeds can in some cases display a non-monotonic dependence on the jump height,
e.g., see \cite{ElGrimshawSmyth2006} for the Serre-Green-Naghdi equations.
In the present case, however, the numerically computed trailing edge speeds  
for both the discrete and continuum model display a monotonic dependence on the size of the jump, 
even though the genuine nonlinearity and strict hyperbolicity of the
modulation equations are presently unknown and
constitute an interesting open problem for future studies.

We believe that this study paves the way for a number of
future explorations in this field, and more generally in the
emergent theme of lattice dispersive hydrodynamics. 
On the one hand, identifying settings (here or in other models)
where quantitative information can be extracted from the
modulation equations would be of considerable and broad interest.
Moreover, developing the systematics of such Whitham modulation
equations at the discrete and at the regularized continuum
settings and comparing the two (and perhaps even different
formulations of the two, e.g., via conservation laws,
Lagrangian formulations etc.) would also be of interest.
However, there also arguably simpler (yet already
complex) and predominantly numerical tasks to also explore.
For instance, here we have focused on cases where $r^{+}$ and
$r^{-}$ are both finite and far away from the $0$ limit. 
The dynamics near vanishing strain and effectively vanishing
precompression displacement are expected to be 
considerably more complex. Moreover, in the present
considerations we have restricted ourselves to simpler
one-dimensional settings. Yet, ongoing, 
recent considerations~\cite{granularBook,magBreathers2D,Leonard11,l11,andrea} have clearly made the case for higher-dimensional
configurations where it is also relevant to consider such DSW structures,
in analogy with corresponding continuum settings,
e.g., in atomic physics~\cite{Hoefer2006}. Such studies are
currently in progress and will be reported in future publications.

\begin{appendix}
\addcontentsline{toc}{section}{Appendix A: Further details on the derivation of the periodic solutions}
\section*{Appendix A: Further details on the derivation of the periodic solutions}
\label{appendixA}
\def\thesection{A}

In this appendix we display the detailed steps of the computation of the periodic wave solutions of the continuum PDE model \eqref{Target PDE model},
specifically, 
Eqs.~\eqref{eq: periodic solution for p = 3/2 -- main body} and \eqref{periodic solution for the case p =2 -- main body}.

\subsection{$p = 3/2$}

For the case of $p = 3/2$, starting from Eq.~\eqref{transformed ODE 2 for p = 3/2}, we can rewrite it as 
\begin{equation}
    \left(g'\right)^{2} = \frac{12}{5\epsilon^{2}c^{2}}\left(g_{1}-g\right)\left(g_{2}-g\right)\left(g_{3}-g\right),
    \label{simplied ODE for p = 3/2}
\end{equation}
where $g_1 < g_2 <g_3$ are the three (suitably defined in connection
to the parameters of the problem) roots of the polynomial $P\left(g\right) = -g^{3} + \frac{5}{4}c^{2}g^{2} - \frac{5}{2}a$.
Separating variables for Eq.~\eqref{simplied ODE for p = 3/2} and integrating both sides with respect to $z$ yields
\begin{equation}
    \int \frac{dg}{\sqrt{\left(g_1 - g\right)\left(g_2 - g\right)\left(g_3 - g\right)}} = \pm \frac{\sqrt{12}}{\sqrt{5}\epsilon c}\left(Z-z_0\right).
    \label{integrating the ODE}
\end{equation}
We then make the familiar (from the case of the KdV) change of dependent variables:
\begin{equation}
    g = g_3 + \left(g_2 - g_3\right)\sin^{2}\theta.
    \label{Change of variables}
\end{equation}
Substituting \eqref{Change of variables} into Eq.~\eqref{integrating the ODE} yields
\begin{equation}
    \int \frac{d\theta}{\sqrt{1 - m\sin^{2}\theta}} = \pm \frac{\sqrt{3\left(g_3 - g_1\right)}}{\sqrt{5}\epsilon c}\left(Z-z_0\right),
    \label{simpled equation}
\end{equation}
where
\begin{equation}
m = \frac{g_3 - g_2}{g_3 - g_1},
\end{equation}
and $z_0$ is a integration constant. 
Using the definition of the Jacobi elliptic function, we rewrite Eq.~\eqref{simpled equation} as follows
\begin{equation}
    \theta = \text{am}\left( \frac{\sqrt{3\left(g_3 - g_1\right)}}{\sqrt{5}\epsilon c}\left(Z-z_0\right), m\right),
\end{equation}
where \text{am} denotes the Jacobi amplitude function.
Now we have that
\begin{align}
    g(Z) &= g_3 + \left(g_2 - g_3\right)\sin^{2}\theta 
    = g_2 + \left(g_3 - g_2\right)\cos^{2}\theta \nonumber \\
    &= g_2 + \left(g_3 -g_2\right)\text{cn}^{2}\left( \frac{\sqrt{3\left(g_3 - g_1\right)}}{\sqrt{5}\epsilon c}\left(Z-z_0\right), m\right)
\end{align}
where $\text{cn}\left(Z,k\right)$ denotes the Jacobi elliptic cosine function.
Finally, since $R = g^{2}$, we solve for $R$ to get the following periodic solution,

\begin{equation}\label{eq: periodic solution for p = 3/2}
    R(Z) = \left[g_{2} + \left(g_{3} - g_{2}\right)\text{cn}^{2}\left( \frac{\sqrt{3\left(g_3 - g_1\right)}}{\sqrt{5}\epsilon c}\left(Z-z_0\right), m\right)\right]^{2}.
\end{equation}

\subsection{$p = 2$}

For the case of $p = 2$, we do not need to apply the transformation $R = g^2$, so we simply focus on the original ODE \eqref{final traveling wave ODE} which now becomes
\begin{equation}
    \frac{\epsilon^{2}c^{2}}{12}\left(R'\right)^{2} = c^{2}R^{2} - \frac{2}{3}R^{3} - 2BR - C.
    \label{ODE for p = 2}
\end{equation}
Separating variables yields
\begin{equation}
    \frac{dR}{\sqrt{-R^{3}+\frac{3}{2}c^{2}R^{2}-3BR-\frac{3}{2}C}} = \pm \frac{\sqrt{8}}{\epsilon c}dZ.
    \label{Simplied ODE for p = 2}
\end{equation}
Then, we denote the three roots of the polynomial $P\left(R\right) = -R^{3}+\frac{3}{2}c^{2}R^{2}-3BR-\frac{3}{2}C$ by $R_1 \leq R_2 \leq R_3$ so that \eqref{Simplied ODE for p = 2} becomes 
\begin{equation}
    \frac{dR}{\sqrt{\left(R_1 - R\right)\left(R_2 - R\right)\left(R_3 - R\right)}} = \pm \frac{\sqrt{8}}{\epsilon c}dZ.
    \label{second simplified ode for p = 2}
\end{equation}
Integrating both sides of Eq.~\eqref{second simplified ode for p = 2} over $Z$ and using the same procedures of computation as in  section 4.1, we obtain
\begin{equation}
    R(Z) = R_{2} + \left(R_{3}-R_{2}\right)\text{cn}^{2}\left( \frac{\sqrt{2\left(R_{3}-R_{1}\right)}}{\epsilon c}\left(Z-z_{0}\right), m\right),
    \label{periodic solution for the case p =2}
\end{equation}
where 
\begin{equation}
m = \frac{R_3 - R_2}{R_3 - R_1}.
\end{equation} 

\end{appendix}

\bibliographystyle{unsrt}

\bibliography{main}

\begin{thebibliography}{10}

\bibitem{scholar}
M.~J. Ablowitz and M.~Hoefer.
\newblock Dispersive shock waves.
\newblock {\em Scholarpedia}, 4(11):5562, 2009.

\bibitem{Mark2016}
G.A. El and M.A. Hoefer.
\newblock Dispersive shock waves and modulation theory.
\newblock {\em Physica D: Nonlinear Phenomena}, 333:11, 2016.

\bibitem{Whitham74}
G.B. Whitham.
\newblock {\em Linear and Nonlinear Waves}.
\newblock Wiley, New York, 1974.

\bibitem{first_DSW}
D.~H. Tsai and C.~W. Beckett.
\newblock Shock wave propagation in cubic lattices.
\newblock {\em J. of Geophys. Res.}, 71(10):2601, 1966.

\bibitem{Nester2001}
V.F. Nesterenko.
\newblock {\em Dynamics of Heterogeneous Materials}.
\newblock Springer-Verlag, New York, 2001.

\bibitem{Hascoet2000}
E.~Hascoet and H.~J. Herrmann.
\newblock Shocks in non-loaded bead chains with impurities.
\newblock {\em Eur. Phys. J. B}, 14:183, 2000.

\bibitem{Herbold07}
E.~B. Herbold and V.~F. Nesterenko.
\newblock Solitary and shock waves in discrete strongly nonlinear double power-law materials.
\newblock {\em Appl. Phys. Lett.}, 90(26):261902, 2007.

\bibitem{shock_trans_granular}
E.~B. Herbold and V.~F. Nesterenko.
\newblock Shock wave structure in a strongly nonlinear lattice with viscous dissipation.
\newblock {\em Phys. Rev. E}, 75:021304, 2007.

\bibitem{Molinari2009}
A.~Molinari and C.~Daraio.
\newblock Stationary shocks in periodic highly nonlinear granular chains.
\newblock {\em Phys. Rev. E}, 80:056602, 2009.

\bibitem{HEC_DSW}
H.~Kim, E.~Kim, C.~Chong, P.~G. Kevrekidis, and J.~Yang.
\newblock Demonstration of dispersive rarefaction shocks in hollow elliptical cylinder chains.
\newblock {\em Phys. Rev. Lett.}, 120:194101, 2018.

\bibitem{granularBook}
C.~Chong and P.~G. Kevrekidis.
\newblock {\em Coherent Structures in Granular Crystals: From Experiment and Modelling to Computation and Mathematical Analysis}.
\newblock Springer, New York, 2018.

\bibitem{yuli_book}
Yu. Starosvetsky, K.R. Jayaprakash, M.~Arif Hasan, and A.F. Vakakis.
\newblock {\em Dynamics and Acoustics of Ordered Granular Media}.
\newblock World Scientific, Singapore, 2017.

\bibitem{gc_review}
C.~Chong, Mason~A. Porter, P.~G. Kevrekidis, and C.~Daraio.
\newblock Nonlinear coherent structures in granular crystals.
\newblock {\em J. Phys.: Condens. Matter}, 29:413003, 2017.

\bibitem{sen08}
S.~Sen, J.~Hong, J.~Bang, E.~Avalos, and R.~Doney.
\newblock Solitary waves in the granular chain.
\newblock {\em Phys. Rep.}, 462:21, 2008.

\bibitem{fleischer2}
Shu Jia, Wenjie Wan, and Jason~W. Fleischer.
\newblock Dispersive shock waves in nonlinear arrays.
\newblock {\em Phys. Rev. Lett.}, 99:223901, Nov 2007.

\bibitem{talcohen}
Jian Li, S~Chockalingam, and Tal Cohen.
\newblock Observation of ultraslow shock waves in a tunable magnetic lattice.
\newblock {\em Phys. Rev. Lett.}, 127:014302, Jun 2021.

\bibitem{Venakides99}
A.~M. Filip and S.~Venakides.
\newblock Existence and modulation of traveling waves in particles chains.
\newblock {\em Comm. Pure and Appl. Math.}, 52(6):693, 1999.

\bibitem{DHM06}
W.~Dreyer, M.~Herrmann, and A.~Mielke.
\newblock Micro-macro transition in the atomic chain via {W}hitham's modulation equation.
\newblock {\em Nonlinearity}, 19(2):471, 2005.

\bibitem{blochkodama}
A.~M. Bloch and Y.~Kodama.
\newblock Dispersive regularization of the whitham equation for the toda lattice.
\newblock {\em SIAM Journal on Applied Mathematics}, 52(4):909--928, 1992.

\bibitem{physd2024v469p134315}
G.~Biondini, C.~Chong, and P.~G. Kevrekidis.
\newblock {On the {Whitham} modulation equations for the {Toda} lattice and the quantitative description of its dispersive shocks}.
\newblock {\em Physica D}, 469:134315, 2024.

\bibitem{marchant2012}
T.~R. Marchant and Noel~F. Smyth.
\newblock Approximate techniques for dispersive shock waves in nonlinear media.
\newblock {\em Journal of Nonlinear Optical Physics and Materials}, 21(03):1250035, 2012.

\bibitem{El2005}
G.~A. El.
\newblock {Resolution of a shock in hyperbolic systems modified by weak dispersion}.
\newblock {\em Chaos: An Interdisciplinary Journal of Nonlinear Science}, 15(3):037103, 10 2005.

\bibitem{Sprenger2024}
Patrick Sprenger, Christopher Chong, Emmanuel Okyere, Michael Herrmann, Panayotis Kevrekidis, and Mark Hoefer.
\newblock Dispersive hydrodynamics of a discrete conservation law.
\newblock {\em arXiv:2404.1675}, 2024.

\bibitem{CHONG2022}
Christopher Chong, Michael Herrmann, and P.G. Kevrekidis.
\newblock Dispersive shock waves in lattices: A dimension reduction approach.
\newblock {\em Physica D: Nonlinear Phenomena}, 442:133533, 2022.

\bibitem{Ari2024}
Christopher Chong, Ari Geisler, Panayotis~G. Kevrekidis, and Gino Biondini.
\newblock Integrable approximations of dispersive shock waves of the granular chain.
\newblock {\em Wave Motion}, 130:103352, 2024.

\bibitem{yasuda}
H.~Yasuda, C.~Chong, J.~Yang, and P.~G. Kevrekidis.
\newblock Emergence of dispersive shocks and rarefaction waves in power-law contact models.
\newblock {\em Phys. Rev. E}, 95:062216, 2017.

\bibitem{Johnson}
K.~L. Johnson.
\newblock {\em Contact Mechanics}.
\newblock Cambridge University Press, Cambridge, UK, 1985.

\bibitem{Yuli2010}
Yuli Starosvetsky and Alexander~F. Vakakis.
\newblock Traveling waves and localized modes in one-dimensional homogeneous granular chains with no precompression.
\newblock {\em Phys. Rev. E}, 82:026603, 2010.

\bibitem{dcdsa}
C.~Chong, P.~G. Kevrekidis, and G.~Schneider.
\newblock Justification of leading order quasicontinuum approximations of strongly nonlinear lattices.
\newblock {\em Disc. Cont. Dyn. Sys. A}, 34:3403, 2014.

\bibitem{Ahnert_2009}
Karsten Ahnert and Arkady Pikovsky.
\newblock Compactons and chaos in strongly nonlinear lattices.
\newblock {\em Physical Review E}, 79(2), February 2009.

\bibitem{rosenau1}
P.~Rosenau.
\newblock Dynamics of nonlinear mass-spring chains near the continuum limit.
\newblock {\em Phys. Lett. A}, 118:222, 1986.

\bibitem{rosenau2}
P.~Rosenau.
\newblock Hamiltonian dynamics of dense chains and lattices: {O}r how to correct the continuum.
\newblock {\em Phys. Lett. A}, 311:39, 2003.

\bibitem{Zabusky}
N.~J. Zabusky and M.~D. Kruskal.
\newblock Interactions of solitons in a collisionless plasma and the recurrence of initial states.
\newblock {\em Phys. Rev. Lett.}, 15:240--243, 1965.

\bibitem{FPUreview}
G.~Gallavotti.
\newblock {\em The Fermi--Pasta--Ulam Problem: A Status Report}.
\newblock Springer-Verlag, Berlin, Germany, 2008.

\bibitem{sulem}
C.~Sulem and P.L. Sulem.
\newblock {\em The nonlinear {Schr\"odinger} equation: self-focusing and wave collapse}.
\newblock Springer, New York, 1999.

\bibitem{HOCHSTRASSER1989259}
D.~Hochstrasser, F.G. Mertens, and H.~Büttner.
\newblock An iterative method for the calculation of narrow solitary excitations on atomic chains.
\newblock {\em Physica D: Nonlinear Phenomena}, 35(1):259--266, 1989.

\bibitem{Chatterjee}
A.~Chatterjee.
\newblock Asymptotic solution for solitary waves in a chain of elastic spheres.
\newblock {\em Phys. Rev. E}, 59:5912, 1999.

\bibitem{pego1}
Jared~M. English and Robert~L. Pego.
\newblock On the solitary wave pulse in a chain of beads.
\newblock {\em Proceedings of the AMS}, 133:1763, 2005.

\bibitem{El_2017}
G.~A. El, M.~A. Hoefer, and M.~Shearer.
\newblock Dispersive and diffusive-dispersive shock waves for nonconvex conservation laws.
\newblock {\em SIAM Review}, 59(1):3–61, January 2017.

\bibitem{strauss2007partial}
Walter~A Strauss.
\newblock {\em Partial differential equations: An introduction}.
\newblock John Wiley \& Sons, 2007.

\bibitem{PRSA283p238}
{G.} Whitham.
\newblock Non-linear dispersive waves.
\newblock {\em Proc. Roy. Soc. Ser. A}, 283:238--261, 1965.

\bibitem{JETP38p291}
A.~V. Gurevich and L.~P. Pitaevskii.
\newblock Nonstationary structure of a collisionless shock wave.
\newblock {\em Zh. Eksp. Teor. Fiz.}, 65:590--605, 1974.

\bibitem{Kamchatnov}
A~M Kamchatnov.
\newblock {\em Nonlinear Periodic Waves and Their Modulations}.
\newblock World Scientific, 2000.

\bibitem{theil}
Florian Theil and Valery~I. Levitas.
\newblock A study of a hamiltonian model for martensitic phase fransformations including microkinetic energy.
\newblock {\em Mathematics and Mechanics of Solids}, 5(3):337--368, 2000.

\bibitem{PhysRevE.88.023016}
N.~K. Lowman and M.~A. Hoefer.
\newblock Dispersive hydrodynamics in viscous fluid conduits.
\newblock {\em Phys. Rev. E}, 88:023016, Aug 2013.

\bibitem{Hoefer2014}
M.A. Hoefer.
\newblock Shock waves in dispersive {Eulerian} fluids.
\newblock {\em J. Nonlinear Sci.}, 24:525–577, 2014.

\bibitem{MielkePatz2017}
Mielke Alexander and Patz Carsten.
\newblock Uniform asymptotic expansions for the fundamental solution of infinite harmonic chains.
\newblock {\em Z. Anal. Anwend.}, 36:437--475, 2017.

\bibitem{ElGrimshawSmyth2006}
G.~A. El, R.~H.~J. Grimshaw, and N.~F. Smyth.
\newblock Unsteady undular bores in fully nonlinear shallow-water theory.
\newblock {\em Physics of Fluids}, 18(2):027104, 02 2006.

\bibitem{Venakides91}
S.~Venakides, D.~Percy, and O.~Roger.
\newblock The {T}oda shock problem.
\newblock {\em Comm. on Pure and Appl. Math.}, 44(8):1171, 1991.

\bibitem{magBreathers2D}
C.~Chong, Y.~Wang, D.~Marechal, E.~G. Charalampidis, Miguel Moler\'on, Alejandro~J. Mart\'inez, Mason~A. Porter, P.~G. Kevrekidis, and Chiara Daraio.
\newblock Nonlinear localized modes in two-dimensional hexagonally-packed magnetic lattices.
\newblock {\em New Journal of Physics}, 23:043008, 2021.

\bibitem{Leonard11}
A.~Leonard, F.~Fraternali, and C.~Daraio.
\newblock Directional wave propagation in a highly nonlinear square packing of spheres.
\newblock {\em Exp. Mech.}, 53:327, 2013.

\bibitem{l11}
A.~Leonard, C.~Daraio, A.~Awasthi, and P.~Geubelle.
\newblock Effects of weak disorder on stress wave anisotropy in centered square nonlinear granular crystals.
\newblock {\em Phys. Rev. E}, 86:031305, 2012.

\bibitem{andrea}
A.~Leonard, C.~Chong, P.~G. Kevrekidis, and C.~Daraio.
\newblock Traveling waves in {2D} hexagonal granular crystal lattices.
\newblock {\em Granular Matter}, 16(4):531, 2014.

\bibitem{Hoefer2006}
M.~A. Hoefer, M.~J. Ablowitz, I.~Coddington, E.~A. Cornell, P.~Engels, and V.~Schweikhard.
\newblock Dispersive and classical shock waves in bose-einstein condensates and gas dynamics.
\newblock {\em Phys. Rev. A}, 74:023623, Aug 2006.

\end{thebibliography}

\end{document}